\documentclass[aps,prc,amsmath,amssymb,twocolumn,preprintnumbers]{revtex4-2}

\usepackage{bm}        
\usepackage{slashed}   
\usepackage{mathtools}
\usepackage{capt-of}

\usepackage{graphicx}  
\usepackage{dcolumn}   
\usepackage{array}
\usepackage{longtable}

\usepackage{tikz}
\usetikzlibrary{decorations.pathreplacing}
\usetikzlibrary{quantikz}

\usepackage{graphicx}
\usepackage{dcolumn}
\usepackage[hidelinks]{hyperref}
\usepackage{color}
\usepackage{hhline}
\usepackage{colortbl}

\newcommand{\black}{\color[rgb]{0,0,0}}

\newcommand{\ketbrac}[2] { {\text{$C_{{#1}{#2}}$}} }
\newcommand{\projector}[2]{{\text{$P^{({#1})}_{{#2}}$}}}
\newcommand{\set}[1]{{\text{$s^{\dagger}_{{#1}}$}}}
\newcommand{\scrap}[1]{{\text{$s_{{#1}}$}}}

\newcommand{\stepS}[1] {{\text{$\mathcal{S}_{{#1}}$}}}
\newcommand{\stepA}[1] {{\text{$\mathcal{A}_{{#1}} $}}}

\newcommand{\innerid} {{\text{$\mathfrak{i}$}}}
\newcommand{\outerid}[1] {{\text{$I^{({#1})}$}}}

\newcommand{\cG}{\mathcal{G}}

\begin{document}
\preprint{Preprint IPARCOS-UCM-26-040,\ \  \ \  ET-0536A-26}

\title{\textbf{Quantum Computers will constrain the Equation of State of Neutron Stars} 
}%

\author{Adri\'an Casta\~no Garc\'{\i}a, Nahia J. Dios Bilbao, J. J. G\'alvez-Viruet$^\dagger$, Felipe J. Llanes-Estrada$^\dagger$ and Mario Logros\'an \'Alvarez.}
 \affiliation{Depto. F\'{\i}sica Te\'orica \& IPARCOS, Univ. Complutense de Madrid, 28040 Madrid, Spain}
 \thanks{$^\dagger$ Contact authors, \tt juagalve@ucm.es, fllanes@fis.ucm.es}

\author{Nicol\'as  M. Arenaza}
\affiliation{Instituto de F\'{\i}sica Te\'orica, IFT-UAM/CSIC, 28049 Madrid, Spain}

\author{Mar\'{\i}a G\'omez-Rocha}%
\affiliation{%
Depto. de F\'{\i}sica At\'omica, Molecular y Nuclear
and Instituto Carlos I de F\'{\i}sica Te\'orica y Computacional,
Universidad de Granada, Granada, Spain
}%
\begin{abstract} 
The Equation of State (EoS) of Nuclear Matter at high densities, and particularly that of neutron stars, resists \textit{ab initio} Quantum Chromodynamics (QCD) computations due to the notorious sign problem of Lattice Gauge Theory at finite chemical potential. A quantum computer deploying QCD in canonical quantization should be able to make substantial progress. We set some basic goals for a future quantum computer to predict the EoS, and thus the basic static observables of the star (mass, radius and Tidal deformability, for example). We then develop the basic theory to address the canonical Hamiltonian in Weyl (time-axial) gauge expressed in normal modes, together with the squared Gauss operator $\mathcal{G}^2$ necessary to execute energy minimization algorithms restricted to the physical Fock subspace. 
Finally, we deploy our particle-quantum register encoding of a generic field theory to demonstrate QCD at finite chemical potential for a few (three-four) particles with a modest number of momentum modes, by simulating the quantum computer on a classical cluster. This opens the  possibility for effective quantum computers to constrain the microscopic physics of neutron stars simultaneously to the operation of third--generation gravitational wave detectors such as the Einstein Telescope, providing more detailed predictions than has been possible until now.\\
\textit{Keywords: Quantum Computing, Dense QCD, Neutron Star EoS, Canonical Hamiltonian, Weyl Gauge, Particle-Register Encoding.}

\end{abstract}

\maketitle
\section{Introduction: Neutron Star Equation of State at Intermediate Densities}
Numerous tests of General Relativity are carried out {\it in vacuo}, around large compact objects, especially black holes. However, the complete study of Einstein's equations $G_{\mu\nu} = \kappa T_{\mu\nu}$ requires microscopic understanding of their right-hand side, the stress-energy tensor, which for a locally isotropic fluid is  ${\rm diag}(\rho,P,P,P)$ in its rest frame, with  $\rho$ and $P$ the energy density and  pressure, respectively. The densest systems available to analyze strong gravity together with matter are Neutron Stars (NSs) and their mergers.
To test new fundamental physics one needs the Equation of State without influence from astrophysical observables, since the effects in want of testing, such as the presence of Dark Matter~\cite{Perez-Garcia:2020lct,Cano:2025isr,Cano:2026yrl,Das:2025fyf,Issifu:2026diw}, or modifications of General Relativity~\cite{NavarroMoreno:2024kkx,Pradhan:2026wte,Bittencourt:2026erx} are embedded in the extraction of those astrophysical observables, potentially causing circular reasoning: we should have methods at hand that allow a clean interpretation of compact-star observations with stress-energy tensors derived from first principles and nuclear-particle physics alone.

\begin{figure}[h]
    \centering
    \includegraphics[width=1\linewidth]{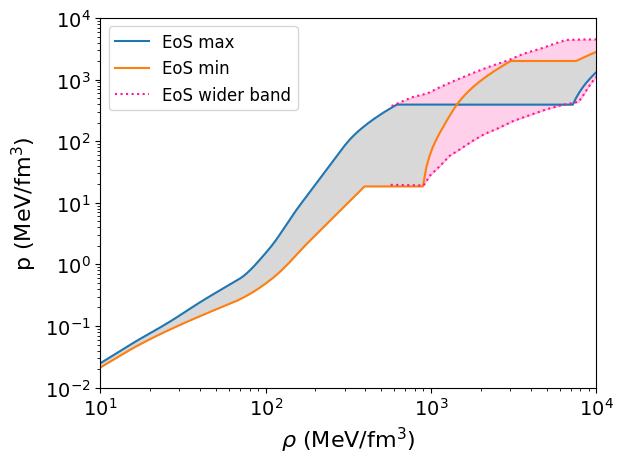}
    \includegraphics[width=1\linewidth]{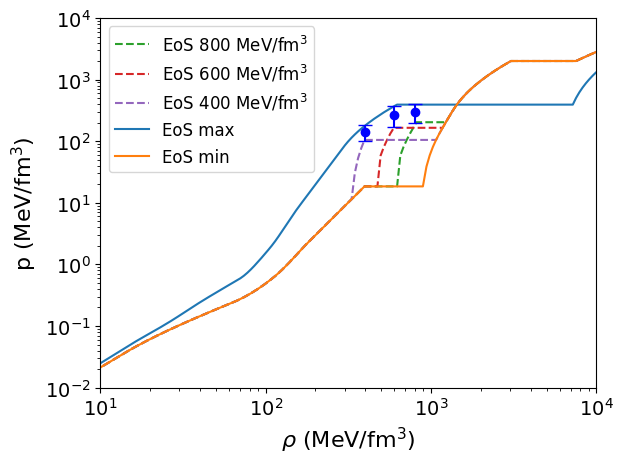}
    \caption{Top panel: Uncertainty band for the EoS $P(\rho)$. \emph{EoS min/max}  are the lower/upper bounds (below an eventual first order phase transition) while respecting
    stability ($c_s\geq0$)/causality ($c_s\leq 1$) once microscopic physics has been accounted , at both low and high $\rho$. Bottom: Eventual improvement of the EoS to narrow the current band, which quantum computer investigations should aim to achieve.}
    \label{fig:EoS}
\end{figure}

A simplified band with the standing constrains~\cite{Alarcon:2024hlj,lope_oter_neos_2019,rhoades_maximum_1974} on the microscopic Equation of State (EoS) $P(\rho)$ of neutron stars can be seen in the top panel of~Fig.~\ref{fig:EoS}.
Nuclear physics, driven for example by density functional theory~\cite{Shchechilin:2024kjv},  extended by chiral perturbation theory (at low density) and perturbative Quantum Chromodynamics, pQCD, (at highest densities) are under relatively good theory control. 

Throughout intermediate densities (and zero temperature), classical simulation techniques for QCD are unworkable because of the sign problem in treating a finite chemical potential in Monte Carlo simulations, and only limited theory progress has been possible~\cite{Cohen:2025ahp}.  In that regime, we are limited to fixing a band determined by stability and causality, $c_s^2\in [0, 1]$,  between the endpoints at the low- and high-density limits where ChPT and perturbative QCD (pQCD) are reliable. (If one samples $P(\rho)$ directly, instead of the thermodynamic potential from which it is derived, then the integral Gibbs-Duhem constraint $\Delta P = \int n(\mu) \,d\mu $ also needs to be imposed~\cite{komoltsev_how_2022}.)
Here is where Quantum Computing could make an important contribution, in constraining $P(\rho)$ around $\rho \sim$ $10^2-10^3$ MeV/fm$^3$, narrowing the uncertainty band and improving the EoS interpolation at intermediate $\rho$. 
We therefore add a very important use case to the increasing menu of applications which quantum computers will have in nuclear and particle physics~\cite{Turro:2024ksf,Barata:2026icn}.

Let us set as a reasonable goal for upcoming quantum computers to obtain one energy density--pressure pair $(\rho, P)$ in the currently inaccessible regime, with an uncertainty at 95\% level \emph{three times smaller} than the presently allowed purely microscopic uncertainty band. This will tighten the allowable EoS which will need to pass through this reduced uncertainty.

We wish to gain a feeling for the progress in the prediction of astrophysical observables within General Relativity that this would bring about, improving the testability of the standard theory of gravity in an extremely dense medium.
For this we solved the Tolman-Oppenheimer-Volkoff (TOV) equations~\cite{tolman_static_1939,oppenheimer_massive_1939} employing the
EoS band's upper and lower limits (dubbed {\it EoS max} and {\it EoS min} respectively) to obtain  corresponding bands for the mass-radius and mass-Tidal deformability diagrams. 

These bands are understood to inherit $2\sigma$ uncertainty from the low-density chiral perturbation theory computation on which they are based, and are valid into the first-order phase transition that both extreme EoS display (after that, the two equations cross in the $(\rho,P)$ plane and a more sophisticated sampling analysis is needed).

To understand where the effect of a single, good-quality, quantum--computed point would be maximum, we have tried three $\rho$ values, $\rho = 400, 600$ and $800$ MeV/fm$^3$ (well above crustal densities~\cite{Burrello:2025jay,Beznogov:2025hrh}). Each has been assigned a pressure with a relative uncertainty of around 30-40\% of the current one and within the current band. The updated EoS are forced to be compatible with this supposed constraint; the bottom panel of Fig.(\ref{fig:EoS}) then shows how the EoS band would  narrow.

The results are promising. From a 36\% relative uncertainty in predicting the stellar radius with current microscopic theory without the quantum computer, we find a reduction down to 15\%. 

For the standard 1.4$M_\odot$ neutron star, the largest gains upon constraining $P(\rho)$ with a quantum computer occur at the lowest of the three employed densities, since the radius is controlled to a larger extent by less dense, outer layers of the neutron star. 

\begin{table}[h]
    \centering
\caption{Radii for a standard 1.4$M_\odot$ star from a first-principles future quantum computer calculation setting alternative constraints to the EoS band
(bottom panel of Fig.\ref{fig:EoS}). The bands are tagged by the first column of the table (the density at which the EoS-band has been artificially constrained).}
\label{tab:tabla R-M}
\begin{tabular}{c||c|c|c}
$\rho$ ($\mathrm{MeV/fm}^3$) & $p$ ($\mathrm{MeV/fm}^3$) & $R$ ($\mathrm{km}$) & Uncertainty (\%) \\ 
\hhline{=#= = = }
800 & $300 \pm 100$ & $12 \pm 3$       & 25 \\ \hhline{-||- - -}
600 & $270 \pm 100$ & $12.5 \pm 2.5$   & 20 \\ \hhline{-||- - -}
400 & $140 \pm 40$  & $13 \pm 2$       & 15 
\end{tabular}

\end{table}

Because Neutron stars are not pointlike objects, in a binary merger they are tidally deformed in the nonuniform field of their companion, leaving an imprint in the outgoing gravitational wave train.
The usual Tidal deformability \eqref{lambda} relates the induced quadrupole to the external tidal perturbation, and stems from the introduction of a quadrupolar correction in the Schwarzschild metric. In geometrodynamic units, $G = c = 1$, $\Lambda$ is a dimensionless parameter. It can be reexpressed~\cite{hinderer_tidal_2010,flanagan_constraining_2008} in terms of the compactness $M/R$ and Love number $k_2$ as
\begin{eqnarray}
    \Lambda = \frac{2}{3}k_2 \frac{R^5}{M^5}\ .
    \label{lambda}
\end{eqnarray}
(Fully relativistic treatments of dynamical tidal responses are also numerically investigated~\cite{Apostolidis:2026qsg} and shall also need uncertainty-controlled microscopic input at the time of third-generation gravitational-wave detectors.)

\begin{table}[h]
 \centering
\caption{ Tidal deformabilities for a 1.4$M_\odot$ star for eventual EoS bands constrained by a quantum computer to pass by  $(\rho, p\pm \Delta p)$ in MeV/fm$^3$. Note that this is obtained from microscopic physics alone, with no astrophysical data input, and thus it would represent the pure-theory prediction.}
\label{tab:tabla Lambda-M}

\begin{tabular}{c||c|c|c}
$\rho\, [\mathrm{MeV/fm}^3]$ & $p \,[\mathrm{MeV/fm}^3]$ & $\Lambda$ & Uncertainty (\%) \\ 
\hhline{=#===}
800 & $300 \pm 100$ & $1050 \pm 950$ & 90 \\ \hhline{-||---}
600 & $270 \pm 100$ & $1100 \pm 900$ & 82 \\ \hhline{-||---}
400 & $140 \pm 40$  & $1225 \pm 750$ & 60
\end{tabular}
     
     \end{table}

\begin{figure}[h]
    \centering
    \includegraphics[width=1\linewidth]{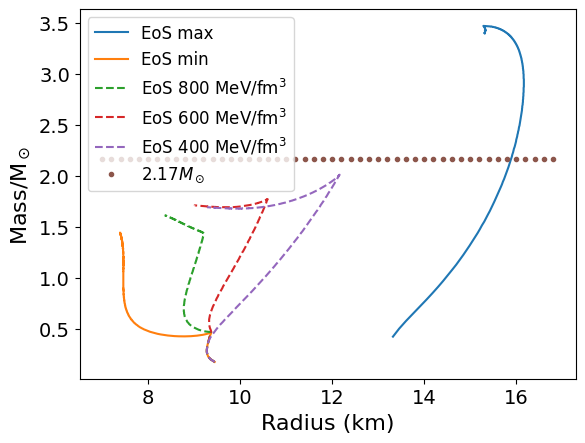}
    \includegraphics[width=1\linewidth]{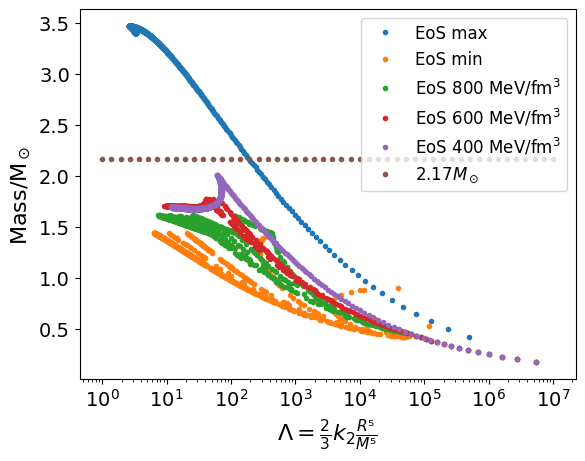}
    \caption{Top panel: Mass (in solar masses) against neutron-star radius. Bottom: Mass against Tidal deformability.  The extreme lines show the current prediction of these observables \emph{excluding astrophysical constraints}, from microscopic theory alone. The intermediate lines, from the left bottom corner towards the right and up, represent the improved predictions if one $(\rho,P)$ point is calculated at the densities marked in the bottom plot of Fig.~\ref{fig:EoS} (note that their uncertainty bands are anchored at the stiffest EoS, so they lift the softest one). }
    \label{fig:Observables}
\end{figure}

From the current~\cite{Alarcon:2024hlj} EoS band (top panel of Fig.~\ref{fig:EoS}), a Tidal deformability of $\Lambda = 1010\pm 990$ is obtained for stars of $1.4M_\odot$, from microscopic physics alone. Narrowing the band as in Fig.(\ref{fig:EoS})-the target we set for a quantum computer-, the uncertainty in this parameter can be reduced (Table \ref{tab:tabla Lambda-M}). The improvement in precision here is not as good as in the radius. However, considering it in logarithmic scale, from the bottom panel of Fig.(\ref{fig:Observables}) we can conclude that results are satisfactory as an initial quantum computer calculation would more clearly exclude $\Lambda=0$. 

We conclude that the lower the density at which the quantum computer can reduce the EoS band, the less uncertainty passed onto the prediction for $\Lambda$ within GR. To have an impact on astrophysical observables, it would be best to gain information on the EoS in the region of densities around 400-500 MeV/fm$^3$. This is no mean feat as QCD is initially formulated as a perturbative theory, here applicable at densities above $5\times 10^3$ MeV/fm$^3$ or so. The quantum computer would have to take this Hamiltonian, formulated in terms of quark--gluon degrees of freedom, and calculate the ground state at densities for which hadrons are likely the active degrees of freedom, thus implementing at least one phase transition (this would not be very surprising given the richness of proposed phases including hyperons~\cite{Kochankovski:2025lqc,Kochankovski:2025thf}, different quark condensates~\cite{Andersen:2026xrf} and more).

The rest of this article is dedicated to setting up the theory and performing a demonstration of how canonically-quantized QCD could be treated on a quantum computer for describing the Neutron-Star Equation of State, a heretofore untractable problem at intermediate densities. 
Eventually, other pieces of the stress-energy tensor such as transport coefficients, which offer particular difficulty to lattice-gauge theory treatments, will also be addressed by the quantum computing community as its involvement in neutron-star matter increases~\cite{Hernandez:2025zxw,Alford:2025tbp}.

\section{Hamiltonian of Chromodynamics in time-axial gauge}
\label{time-axialH}


For static systems that entail a fixed reference frame, such as neutron star cores, Lorentz covariance can be traded for computational simplicity by adopting a frame-dependent formalism. As we take a fixed gauge, we do not pursue a lattice quark-link Hamiltonian approach but a canonical quark-gluon one. Also, although the overall structure of the star requires a General Relativistic treatment, the microscopic physics happens in a span too short to see any curvature (fm {\it vs.} km), so that all derivatives in what follows are simple partial ones.

In the time-axial gauge the temporal component of the gluon field is set to zero, $A^0(\vec{x},t_0)=0$. The spatial components are then considered independent and fulfill the canonical commutation relations:
\begin{equation}
\left[A^i(\vec{x},t_0),A^j(\vec{y},t_0)\right] = \delta^{ij}\,\delta^{(3)}(\vec{x}-\vec{y})\ .
\end{equation}
It is now possible to define generalized momenta 
\begin{equation}
    \Pi^i = \frac{\delta L_{QCD}}{\delta(\partial_0A^i)}\ , \ \ \  \ \ \ \ \ \pi =\frac{\delta L_{QCD}}{\delta(\partial_0\psi)}\ ,
\end{equation}
and after  a Legendre transformation~\cite{lee_particle_1988}, the time-axial gauge QCD Hamiltonian is particularly simple:
\begin{align}
& H_{QCD} =  \sum\Pi^i \dot{A}^i + \pi \dot{\psi} -L_{QCD} \nonumber\\
    & =\int d^3x \left\{
\tfrac{1}{2}\left[(E_i^a)^2 + (B_i^a)^2\right] + g\, \vec{A}^a \cdot (\psi^\dagger_l \gamma^0 \vec{\gamma} T^a \psi_l)
\right. \notag\\
&\qquad \left.
+ \psi_l^\dagger \left( -i \gamma^0 \vec{\gamma} \cdot \vec{\nabla} + m \gamma^0 \right) \psi_l
\right\}\ .
\label{HQCD}
\end{align}
Therein, $E_i^a$ and $B_i^a$ are the chromo-electric and chromo-magnetic fields, defined as $E^i = -G^{0i}$ and $B^i = -\epsilon^{ijk}G^{jk}$, with
\begin{equation}
    G_{\mu \nu }= \partial_\mu A_\nu -\partial_\nu A_\mu +ig[A_\mu, A_\nu]\ .
\label{G_munu}
\end{equation}
The difficulty with the $A^0=0$ choice is that it does not fix the gauge completely, and must be complemented with a subsidiary equation over the Hilbert space to guarantee Gauss's law, to name it 
\begin{equation}
    \mathcal{G}^a (\vec{x}, t_0) \ket{\Psi} = 0\ ,
\end{equation}
where $\mathcal{G}$ is the Gauss' law operator, derived from the Lagrangian density $\mathcal{L}_{QCD}$ as the would-be Euler-Lagrange equation, which in components reads
\begin{align}
\mathcal{G}^a & = \partial_\mu\frac{\partial\mathcal{L}_{QCD}}{\partial(\partial_\mu A^{0\ a})}-\frac{\partial\mathcal{L}_{QCD}}{\partial A^{0\ a}}\nonumber\\
& = \left(\delta^{ab}\nabla_i + gf^{abc}A^{c\ i}\right)\,E^{b\ i} +g\,\psi^{\dagger}T^a\psi\ .
\end{align}
Some further details about this operator can be found in section \ref{Apendix:Gauss} of the appendix.

\subsection{Normal Mode Expansion}
For an investigation on the Equation of State, a property of equilibrium, we aim to minimize the expectation value of the Hamiltonian at fixed time, so we drop explicit reference to time, i.e. $A^{i,a}(x) = A^{i,a}(\vec{x},t_0)$, etc. and expand the fields in a plane wave basis of definite momentum states, introducing creation and annihilation operators.  Gluon fields are
\begin{align}
    \vec{A}^a (x) &= \int [k]\sum_{\sigma = 1}^3 \vec{\epsilon}_{\sigma }[a_{k\sigma a}\,e^{ik·x}+a_{k\sigma a}^\dagger \,e^{-ik·x}]
    \label{A(a+,a)}\ , \\
    \vec{E}^a (x) &= i\int [k]\sum_{\sigma = 1}^3 k^0\vec{\epsilon}_{\sigma }[a_{k\sigma a}\,e^{ik·x}-a_{k\sigma a}^\dagger\, e^{-ik·x}]
    \label{E(a+,a)}\ ,
\end{align}
where $\int[k] = \int d^3k /\left((2\pi)^{3/2}\sqrt{2E_k}\right)$ and $\vec{\epsilon}_\sigma$ is a Cartesian polarization vector chosen by the subindex $\sigma$,
\begin{equation}
\vec{\epsilon}_1 = \hat{\imath}, \,\,\, \vec{\epsilon}_2 = \hat{\jmath}, \,\,\,\vec{\epsilon}_3 = \hat{k}\ .
\label{def:gluon-pol}
\end{equation}
The quark fields are
\begin{equation}
\psi_{cf}(\vec{x}) =\int [k] \left( e^{ikx}\,u_{\lambda f}(k)\, b_{k\lambda cf}\,+e^{-ikx}\,v_{\lambda f}(k)\, d^\dagger_{k\lambda c f}\right),
    \label{Dirac spinor}
\end{equation}
where the spinor basis can be found in section \ref{sec:spinor_basis} of the appendix. 
The creation and annihilation operators fulfill the canonical commutation relations
\begin{align}
\left[a_{k\sigma a},a^{\dagger}_{p\lambda b}\right] &= \delta_{ab}\,\delta_{\sigma\lambda}\,\delta^{(3)}(\vec{p}-\vec{k})\ ,\label{bosonscommute} \\
\left\{b_{k\sigma cf},b^{\dagger}_{p\lambda bc'f'}\right\} &= \left\{d_{k\sigma cf},d^{\dagger}_{p\lambda c'f'}\right\} \nonumber\\ 
& = \delta_{ff'}\delta_{cc'}\,\delta_{\sigma\lambda}\,\delta^{(3)}(\vec{p}-\vec{k})\ , \label{fermionsanticommute}
\end{align}
with all other (anti-)commutators zero. The complete expansion of the Weyl-gauge QCD Hamiltonian of Eq.~(\ref{HQCD}) in terms of these creation and annihilation operators can be found in section \ref{Apendix:Hamiltonian-expansion} of the appendix. 

\subsection{Discretization}
The Hamiltonian Eq.~(\ref{HQCD}) is discretized on a three-dimensional rectangular domain with periodic boundary conditions. We introduce an ultraviolet cutoff $\Lambda$ (not to be confused with the Tidal deformability of Eq.~(\ref{lambda})) and integer indices $m_i$ and $n_i$ for each dimension 
\begin{equation}
    p_i = \frac{\Lambda}{1+2N_i}\,m_i\ ,\ \  x_i = \frac{2\pi}{\Lambda}n_i\ ,
    \label{def:momentum-x}
\end{equation}
with 
\begin{equation}
m_i, n_i \in \left\{-N_i, ...,0,...,N_i\right\}\ ,
\label{integerslabels}    
\end{equation}
and $i$ going from 1 to 3. 
With this set of indices,
\begin{align}
\prod_i\sum_{p_i} e^{-i\,p_i (x_i-y_i)} &= \prod_i\sum^{N_i}_{m_i =-N_i}e^{-i\,2\pi \frac{m_i (n_i-n'_i)}{1+2N_i}} \nonumber\\
& = \prod_i (2N_i+1)\,\delta_{n_i n'_i}\ .
\label{sumexps}
\end{align}
The volume is then related to the ultraviolet cutoff by
\begin{equation}
V = \frac{(2\pi)^3(1+2N_1)(1+2N_2)(1+2N_3)}{\Lambda^3}.
\label{eq:volume}
\end{equation}

The zero element in Eq.~(\ref{integerslabels})  will be excluded from the integration grid as a way of regulating the infrared. This is the same as adopting the infrared cutoffs $\lambda_i=\frac{\Lambda}{1+2N_i}$, and it has the effect of removing off  Eq.~(\ref{sumexps}) the unit from $(2N_i+1)$, which is a small discretization error for sufficiently sizeable $N_i$.
The integral measure and the creation and annihilation operators are discretized as
\begin{equation}
\int d^3k\rightarrow \sum_{\vec{m}}\left(\frac{\Lambda}{1+2N}\right)^3,\ a_{\vec{k}} \rightarrow a_{\vec{m}}\left(\frac{1+2N}{\Lambda}\right)^{3/2}\ ,
\end{equation}
where we use the shorthand $ \prod^3_{i=1} (1+2N_i)\to (1+2N)^3 $.
In the following, momenta are only taken along the $\hat{\imath}$ direction 
so we will drop the direction subindex and write $N$ instead of $N_1$. 
\subsection{Encoding}
\label{subsec:encoding}
The Particle-Register Encoding (PRE) of creation and annihilation operators developed at Complutense has been introduced and extensively described in \cite{galvez-viruet_dynamical_2024}, and we will only very briefly recall it here.
It enables both the construction of wavefunctions (wf) and the Pauli string decomposition of operators in a unified framework. 
The motivation for such computing method is that it puts at our disposal a dynamic memory handling representing, in an intuitive way, the particle creation-destruction intrinsic to field theory, fully entangling states with varying particle number.

In short, quark and gluon states are encoded in computational basis states on qubit registers.
We denote by $\left|v_{q\,(g)}\right\rangle$ a register which could contain one fermion (or gluon) but is momentarily empty,
\begin{equation}
\left|v_{q(g)}\right\rangle = \left|0\right\rangle_{\mathcal{P}}\,\bigotimes_\alpha\,\left|0\right\rangle^{\otimes n^{q\,(g)}_\alpha}\ .
\label{qg-regs}
\end{equation}
Here,   $\ket{0},\ket{1}$ are the single-qubit computational basis states, 
$\alpha$ the possible quantum numbers (color, spin/polarization, momentum, eventually flavor) of the particle which can occupy the register,
and $n^{q\,(g)}_\alpha$ the number of qubits needed to encode each $\alpha$. The symbol $\mathcal{P}$ tags a special \emph{presence} qubit controlling whether the register currently encodes the vacuum (is empty) or a single-particle configurations (is occupied); the remaining qubits are collectively denoted as $\textit{mode}$ qubits. The vacuum appears when all register qubits are in the $\ket{0}$ state, as opposed to the single-particle states, which only require the presence qubit to take the value $\ket{1}$. To pass from/to  empty to/from occupied registers  we use custom $s^{\dagger}$ ($\textit{set}$) and $s$ ($\textit{scrap}$) operators:
\begin{align}
\ket{(p,i,a)} & = C_{10}\otimes s^{\dagger}_{pia}\left|v_{g}\right\rangle \nonumber\\
& = C_{10}\ket{0}_{\mathcal{P}}\otimes s^{\dagger}_{p}\ket{0}^{\otimes n^g_m}\otimes s^{\dagger}_{i}\ket{0}^{\otimes n^g_s}\otimes s^{\dagger}_{a}\ket{0}^{\otimes n^g_c}\nonumber\\
&  =\ket{1}_{\mathcal{P}}\otimes \ket{p}\otimes \ket{i}\otimes\ket{a}\nonumber\\
& = \ket{1pia},
\label{def:single-particle-set}
\end{align}
where 
\begin{equation}
    C_{ij} \equiv \ket{i}\bra{j},
    \label{def:Cij}
\end{equation}
acts on the presence qubit and $n^{g}_{m},n^{g}_{s},n^{g}_{c}$ are the number of qubits in which the momentum, spin and color states are encoded. They are given by 
\begin{align}
n^q_m = \lceil\log_2  2N\rceil,\quad &n^g_m = \lceil\log_2  2N\rceil, \\
n^q_c = \lceil\log_2  N_c\rceil,\quad & n^g_c = \lceil\log_2  (N_c^2-1)\rceil,\\
n^q_s = \lceil\log_2  N^q_s\rceil,\quad & n^g_s = \lceil\log_2  N^g_s\rceil,\\
n^q_f = \lceil\log_2  N_f\rceil,\quad &
\end{align}
where $2N$ is the number of points in the momentum grid, $N_c$ and $N_f$, 
the numbers of colors and flavours,
$N^q_s$  and $N^g_s$ those of quark spins and gluon polarizations. 

Similar set and scrap operators can be defined to create and annihilate single-quark states, with the addition of $n^q_f$ qubits for the flavor quantum number. For notational simplicity the collective label $(p,i,a)$ will be replaced by a single symbol from now on, so we will write $s^{\dagger}_p$ instead of $s^{\dagger}_{pia}$. 

The tensor product of registers allows for the representation of multi-particle states. To encode up to $n^q$ quarks and $n^g$ gluons we take
\begin{equation}
   \ket{v} = \ket{v_q}^{\otimes n^q}\otimes\ket{v_g}^{\otimes n^g}\ ,
   \label{eq:total-vac}
\end{equation}
states with $i\leq n^q$ ($j\leq n^g$) quarks (gluons) are encoded in the rightmost $i$ ($j$) quark (gluon) registers. To maintain the filling ordering we define projectors over states with $i$ particles:
\begin{align}
P^{(n^q)}_i &\ket{v_q}^{\otimes n^q-j}\otimes\ket{1p_j}\dots\ket{1p_1} \nonumber \\
    &= \delta^i_j\ket{v_q}^{\otimes n^q-j}\otimes\ket{1p_j}\dots\ket{1p_1}\ ,
    \label{def:projectors}
\end{align}
and similarly for gluon registers. More explicitly
\begin{equation}
P^{(n^q)}_j \equiv (C_{00}\otimes\innerid)^{\otimes(n_q-j)}\otimes (C_{11}\otimes\innerid)^{\otimes j} ,
\label{eq:projector-explicit}
\end{equation}
where $\innerid$ is the identity over mode qubits.
Combining the $\textit{set}$ and $\textit{scrap}$ operators of Eq.~(\ref{def:single-particle-set}) with the projectors of Eq.~(\ref{def:projectors}) we define the PRE of creation and annihilation operators of quarks and gluons
\begin{align}
b^{(n^q)\dagger }_{p} &=  \sum^{n^q}_{j = 1}\stepA{j}\cdot \projector{n^q-j}{0}\otimes \left(\ketbrac{1}{0}\otimes \set{p}\right)_{j}
\otimes \,\projector{j-1}{j-1}, \label{def:nReg-fermioncreation}\\
a^{(n^g)\dagger }_{p} & =  \sum^{n^g}_{j = 1}  \stepS{j} \cdot \projector{n^g-j}{0}\otimes \left(\ketbrac{1}{0}\otimes \set{p}\right)_{j} \otimes \,\projector{j-1}{j-1},
\label{def:nReg-bosoncreation}
\end{align}
where $\stepS{j}$, $\stepA{j}$ are step-symmetrizer and antisymmetrizer of the state of the $j$th register with all the previous ones (counting from right to left):
\begin{equation}
    \stepS{j} \equiv \frac{1}{\sqrt{j}}\left(\outerid{n} + \sum^{j-1}_{i=1}\mathcal{P}_{ji}\right),\, \stepA{j} \equiv \frac{1}{\sqrt{j}}\left(\outerid{n}-\sum^{j-1}_{i=1}\mathcal{P}_{ji}\right),
\label{def:step-symmetrizer}
\end{equation}
$\mathcal{P}_{ij}$ being an operator that permutes particles $i$ and $j$. Equations~(\ref{def:nReg-fermioncreation}) and (\ref{def:nReg-bosoncreation}) fulfill the canonical (anti-) commutation relations of Eq.~(\ref{bosonscommute}--\ref{fermionsanticommute}) up to a boundary term, see section 4 and 5 and appendix B of \cite{galvez-viruet_dynamical_2024}.

\section{From Hamiltonian to EoS}
The partition function of an isolated system with a non-zero quark chemical potential is
\begin{equation}
Z\,(T,\mu) = \text{Tr}\left\{\exp\,\left[-\beta\,(H-\mu_f N_f)\right]\right\}
\end{equation}
where $H$ is the Hamiltonian of Eq.~(\ref{HQCD}), $N_f$ a quark-number operator with flavor $f$ and $\mu_f$ its  chemical potential. A sum over quark flavors is implicit. 

For a ``low''-temperature system such as the core of a neutron star, the thermal distribution is dominated by the state $\ket{\Psi(\theta)}$ minimizing
\begin{equation}
F(\theta) = \bra{\Psi(\theta)} \,(H-\mu_f N_f)\,\ket{\Psi(\theta)}\,.
\label{cost-function}
\end{equation}
over the space of states satisfying Gauss's law
(to respect the overall gauge invariance of the theory),
\begin{equation}
    \mathcal{G}^a(x)\,\ket{\bar{\Psi}} = 0
    \label{eq:gauss-law}
\end{equation}
for every $a$ and $x$. 

Since $\mathcal{G}^a=\left(\mathcal{G}^a\right)^\dagger$ is Hermitian, its square is nonnegative as we now exploit. First, note that
\begin{equation}
    \bra{\bar{\Psi}}\mathcal{G}^a(x)\mathcal{G}^a(x)\ket{\bar{\Psi}} =\bra{\bar{\Psi}}\mathcal{G}^a(x)^\dagger\mathcal{G}^a(x)\ket{\bar{\Psi}}=0
\end{equation}
and introducing the completeness relation,
\begin{equation}
    \sum_{\phi}\bra{{\Psi}}\mathcal{G}^a(x)^\dagger\ket{\phi} \bra{\phi}\mathcal{G}^a(x)\ket{{\Psi}}=\sum_\phi\left|\lambda^a_\phi(x)\right|^2
    \geq0 \,,
\end{equation}
each piece in the sum has to be separately positive or null:
it is clear that $\langle\mathcal{G}^a(x) \mathcal{G}^a(x)\rangle\geq0$, so the equation
\begin{equation}
    \sum_a\int d^3x\ \langle\mathcal{G}^a(x)\mathcal{G}^a(x)\rangle=
    \bigg\langle\sum_a\int d^3x\ \mathcal{G}^a(x)\mathcal{G}^a(x)\bigg\rangle=0
\end{equation}
requires each term in the sum to be exactly $0$.

Thus, defining the squared Gauss' law operator
\begin{equation}
    \mathcal{G}^2=\sum_{a}\int d^3x\ \mathcal{G}^a(x)\mathcal{G}^a(x) \,,
\end{equation}
Eq.~(\ref{eq:gauss-law}) can be implemented in a variational spirit by just minimizing $\bra{\bar{\Psi}}\mathcal{G}^2\ket{\bar{\Psi}}=0$.
To achieve the simultaneous minimization of $\langle H \rangle$ and $\langle \mathcal{G}^2\rangle$ we optimize $\langle H \rangle +\lambda \langle \mathcal{G}^2 \rangle$ (essentially, Lagrange's multiplier method for a constrained Hamiltonian).

The energy, pressure and sound velocity are then obtained from the expectation value of the Hamiltonian $ \bar{E}=\bra{\bar{\Psi}}H\ket{\bar{\Psi}}$:
\begin{equation}
    \rho = \frac{\bar{E}}{V}\,,\ \ p=-\left.\frac{\partial \bar{E}}{\partial V}\right|_{\mu}, \ \ c^2_s = \frac{\partial p}{\partial\rho},
    \label{def:pres-energydensity}
\end{equation}
where $V$ corresponds to the volume of small and independent cells (with cross-cell interactions negligibly small). Since density at the core of neutron stars is believed to reach about 6–7 times nuclear saturation density, the relevant degrees of freedom are expected to remain predominantly hadronic. In this regime, QCD partonic correlations are expected to be effectively localized over length scales of the size of a few nucleons, i.e. a few femtometers. In fact, since the chemical potential suppresses antiquark excitations, the cell sizes should be around the spatial scales associated with nucleon resonances such as the $\Delta$ and $N^{*}$ \cite{fernandez_quark_2020}, the main {\it in medio} excitation channels.

Realizing such a phenomenologically motivated description would require at least six ``constituent'' quarks, together with the gluonic degrees of freedom necessary to satisfy Gauss’ law. As we shall see, systems of these sizes remain beyond the capabilities of current classical and quantum computers. The proof-of-concept variational wavefunctions considered in the present work are substantially smaller: configurations with up to three quarks and one gluon in a color singlet, sufficient to introduce dynamics into the quark-based description of the nucleon, and configurations with up to one quark and two gluons in a color octect, which allow for the examination of the effect of gluon self-interactions. The present work should therefore be regarded strictly as a proof-of-concept study rather than a realistic description of dense QCD matter.
\black
\section{Illustrative computation}

To minimize the cost function we use the VQE \cite{peruzzo_variational_2014,tilly_variational_2022}, a variational algorithm that combines the use of quantum hardware to represent wavefunctions and measure  expectation values with a classical update of parameters. The classical minimization algorithm we used is the BFGS \cite{nocedal_numerical_2006}, a gradient-based method that iteratively construct the Hessian matrix of the cost function and ensures that it stays positive definite by satisfying the strong Wolfe conditions in each iterative step.  

Gauss' law is imposed in runtime by means of the $\textit{quadratic penalty method}$, which consists of adding the square of the constraint to the cost function,
\begin{equation}
F(\theta)\rightarrow F_1(\theta) = F(\theta)+\frac{\tilde{u}}{\Lambda^2}\,(\mathcal{G}^2(\theta)-\mathcal{G}_{\text{min}}^2)\ ,
\label{def:CF1}
\end{equation} 
and repeatedly running the minimization algorithms with increasing values of $\tilde{u}$ until convergence. It can be shown that, in the limit where $\tilde{u}\rightarrow \infty$, the solution satisfies the constraint while minimizing the cost function $F(\theta)$, see chapter 17 of  \cite{nocedal_numerical_2006}. In our case, convergence is fairly quick if the minimization does not get trapped in a local minimum; typically about three runs are sufficient, with $\tilde{u}$ increasing from 1 to 40. The factor $\mathcal{G}^2_{min}$ is found by minimizing the Gauss' law operator alone and 
it only asymptotically decreases towards zero.

In the remainder of this section, we describe the construction of variational wavefunctions and the decomposition of the Hamiltonian operator in terms of Pauli strings (a common strategy for measuring expectation values on quantum computers). 

\subsection{Variational wavefunctions}
\label{sec:varwaves}
The number of classically simulatable qubits on a personal computer or small cluster is limited to roughly thirty, as storing the corresponding wavefunctions in double-precision complex format already requires roughly 128 GB of memory. The maximum number of particles of the ansatz wavefunctions we are going to use therefore depends on the qubit requirements for each particle register, which are summarized in Tables~\ref{tab:encoding} and~\ref{tab:discretization}. Including the presence qubit, each quark requires a total of seven qubits and each gluon a total of eight, even for the tiny number of available momenta. Thus, wavefunctions are presently limited to contain up to four particles.

Before describing the variational wavefunctions in detail, a few remarks concerning Tables~\ref{tab:encoding} and~\ref{tab:discretization} are in order; momentum indices are given physical dimensions of GeV by multiplication with the discretization factor of Eq.~(\ref{def:momentum-x}) and are taken along the $\hat{\imath}$ direction, furthermore, flavor indices correspond to $u$ and $d$ (strangeness~\cite{Passarella:2026mxy,Becerra:2025ryy} is not included for the time being). 
Note that the limited range of quantum-number values is a consequence of the restricted number of qubits. Including additional values would be straightforward given sufficient resources, see section \ref{scaling} for a discussion about scaling. 
While the number of particles is very small, one should note that they are studied on a finite density bath characterized by a chemical potential $\mu$.
\setlength{\tabcolsep}{5pt}
\renewcommand{\arraystretch}{1.5}

\begin{table}[]
\centering
\caption{Quark quantum-number values and qubit states in which they are encoded. Momenta are given physical dimensions of GeV by multiplication with the discretization factor of Eq.~(\ref{def:momentum-x}).\label{tab:encoding}}
\begin{tabular}{c||c|c|c}
Quarks   & Values      & Binary encoding                                              & Qubits \\ 
\hhline{=#= = = }
Momentum & $1,-1,2,-2$ & $\ket{00},\ket{01},\ket{10},\ket{11}$ & 2      \\  \hhline{-||- - -}
Spin     & $-1,1$      & $\ket{0},\ket{1}$                                         & 1      \\ 
\hhline{-||- - -}
Color    & $1,2,3$     & $\ket{00},\ket{01},\ket{10}$                              & 2      \\ 
\hhline{-||- - -}
Flavor   & $1,2$       & $\ket{0},\ket{1}$                                         & 1     
\end{tabular}
\end{table}

\begin{table}[]
\caption{Gluon quantum-numbers values and qubit states in which they are encoded.}
\begin{tabular}{c||c|c|c}
Gluons   & Values      & Binary encoding                                                  & Qubits \\ 
\hhline{=#= = = }
Momentum & $1,-1,2,-2$ & $\ket{00},\ket{01},\ket{10},\ket{11}$ & 2      \\  \hhline{-||- - -}
Polarization     & $1,2,3$      & $\ket{00},\ket{01},\ket{10}$                                         & 2      \\ 
\hhline{-||- - -}
Color    & $1,...,8$     & $\ket{000},...,\ket{111}$                              & 3       
\end{tabular}

\label{tab:discretization}
\end{table}

\subsubsection{3 Quarks + 0, 1 Gluon.}
\label{vwave:3q1g}
The first wavefunction is a superposition of states with three quarks and three quarks and a gluon:
\begin{equation}
    \ket{\Psi_1(\theta)} = \cos(\theta_1)\ket{\Psi^{3q}} +  \sin(\theta_1)\,e^{i\theta_2}\ket{\Psi^{3q1g}}\ ,
    \label{wavefunction1}
\end{equation}
where $\theta_1,\theta_2$ are the first two parameters. Note that the only phase introduced via parameters is that of Eq.~(\ref{wavefunction1}), all other amplitudes are written in terms of real trigonometric functions.

By commuting the tensor product of Hilbert spaces which form each register,
we can describe a given fixed-particle number sector 
by a product Hilbert space, 
\begin{equation}
    \mathcal{H}^{3q} = \mathcal{H}^{3q}_m\otimes\mathcal{H}^{3q}_s\otimes\mathcal{H}^{3q}_c\otimes\mathcal{H}^{3q}_f\ ,
\end{equation}
where $m$ denotes momentum, $s$ spin, $c$ color and $f$ flavor. The superscript $(3q)$  is changed accordingly to make reference to the sector with three quarks and one gluon. 
The three-quark Fock sector is therefore generated by
\begin{equation}
    \ket{\Psi^{3q}} \in
\mathrm{span}\left\{ \big|m^{3q}_i\big\rangle\otimes\big|s^{3q}_j\big\rangle\otimes\big|c^{3q}_k\big\rangle\otimes\big|f^{3q}_{l}\big\rangle\right\}\ ,
\end{equation}
where $\big|m^{3q}_i\big\rangle,\big|s^{3q}_j\big\rangle,\big|c^{3q}_k\big\rangle,\big|f^{3q}_l\big\rangle$ are independent basis vectors for each quantum number obtained from combinations of values in Tables~\ref{tab:encoding} and~\ref{tab:discretization}. Several constraints are imposed so that the variational wavefunction has enough freedom to resemble that of a neutron. The overall state $\ket{\Psi_1(\theta)}$ is required to be a color singlet, the quark spin and flavor wavefunctions are mixed-symmetric under particle exchange, the momentum wavefunction is symmetric under particle exchange and the total momentum in each Fock sector vanishes.

These constraints are applied to both the three-quark and three-quark one-gluon sectors. In the case of the three-quark component, there are only two independent momentum-basis vectors which are symmetric under particle exchange and sum up to zero. Furthermore, since the color-singlet state is antisymmetric, the spin-flavor component must be symmetric under quark exchange. As the flavor wavefunction transforms in a mixed-symmetry representation, it must be combined with mixed-symmetry spin states. This leaves only two independent spin-flavor basis vectors. Altogether, the three-quark component yields four independent basis amplitudes, see Table~\ref{tab:3q-symmetry}.

The three-quark plus one-gluon component is more involved. To obtain an overall color singlet, the color-octet representation of the three-quark subsystem is combined with the gluon octet. The resulting three-quark color structure is mixed symmetric under quark exchange, and must be combined with the mixed-symmetry flavor and spin parts, yielding two independent basis vectors for the flavor-spin-color sector. Combining these with the three independent gluon polarization states and the eight independent symmetric momentum states gives a total of thirteen amplitudes, see Table~\ref{tab:3q1g-symmetry}.

Overall, the 17 real and 2 complex amplitudes of the total wavefunction are specified by 15 parameters when normalization is taken into account. All amplitudes are real with the exception of the amplitude of $\ket{\Psi^{3q1g}}$ in Eq.~(\ref{wavefunction1}).
\begin{table}[]
\caption{Symmetry under quark exchange and number of independent parameters for the three-quark sector, $S$, $M$ and $A$ stand for symmetric, mixed and antisymmetric, respectively.}
\label{tab:3q-symmetry}
\begin{tabular}{c||cccc}
$3q$ Sector & Momenta & Spin & Flavor & Color \\ 
\hhline{=#====}
\begin{tabular}[c]{@{}c@{}}Sector \\ Symmetry\end{tabular} & S & M & M & A \\ 
\hhline{-||----}
\begin{tabular}[c]{@{}c@{}}Combined \\ Symmetry\end{tabular} & S & \multicolumn{2}{c}{S} & A \\ 
\hhline{-||----}
Amplitudes & 2 & \multicolumn{2}{c}{2} & $\times 1$
\end{tabular}
\end{table}

\begin{table}[]
\caption{Symmetry under quark exchange and number of amplitudes for the three-quark plus one gluon sector. Adding the other three amplitudes that specify the polarization state of the gluon gives a total of 13 amplitudes.}
\label{tab:3q1g-symmetry}
\begin{tabular}{c||cccc}
$3q1g$ Sector & Momenta & Spin & Flavor & Color \\ 
\hhline{=#====}
\begin{tabular}[c]{@{}c@{}}Sector\\ Symmetry\end{tabular} & S & M & M & M \\ 
\hhline{-||----}
\begin{tabular}[c]{@{}c@{}}Combined \\ Symmetry\end{tabular} & S & \multicolumn{3}{c}{A} \\ 
\hhline{-||----}
Amplitudes & 8 & \multicolumn{3}{c}{2}
\end{tabular}
\end{table}

\subsubsection{1 Quark + 0, 1 or 2 Gluons}
\label{vwave:1q2g} Restricting the Fock space to at most one gluon 
closes the exploration of the non-Abelian interaction terms appearing in both the Hamiltonian and the squared Gauss' law operator. To probe the effect of these interaction terms, we considered states with one quark and up to two gluons:
\begin{align}
\ket{\Psi_2(\theta)} = &\cos(\theta_1)\ket{\Psi^{1q}} +  \sin(\theta_1)\cos(\theta_3)\,e^{i\theta_2}\ket{\Psi^{1q1g}} \nonumber \\
&  +\sin(\theta_1)\sin(\theta_3)\,e^{i\theta_4}\ket{\Psi^{1q2g}}\ , \,
\label{wavefunction2}
\end{align}
Similarly to the previous case, complex amplitudes are introduced only between different Fock sectors. Color, in contrast to subsec.~\ref{vwave:3q1g}, belongs to the triplet representation. The cardinal of the independent wavefunction set can be obtained from table~\ref{tab:1q2g-symmetry}. Additionally, choosing the center of mass frame with a vanishing total momentum is imposed only at the level of expectation values, separately in each $n$-particle sector:
\begin{align}
0  = \bra{\Psi^{1q}}\hat{P}\ket{\Psi^{1q}} & = \bra{\Psi^{1q1g}}\hat{P}\ket{\Psi^{1q1g}} \nonumber \\
& = \bra{\Psi^{1q2g}}\hat{P}\ket{\Psi^{1q2g}}\ ,
\end{align}
where $\hat{P}$ is the momentum operator. Such restriction is imposed by construction in the $1q$ sector only, while for the components with one and two gluons we again apply the quadratic penalty method by adding to the cost function the sum of the squares of the momentum expectation values:
\begin{equation}
F_1(\theta)\rightarrow F_2(\theta) = F_1(\theta) + \frac{\tilde{u}}{\Lambda}\sum_{\alpha}\left(\bra{\Psi^{\alpha}(\theta)}\hat{P}\ket{\Psi^{\alpha}(\theta)}\right)^2,
\label{def:CF2}
\end{equation}
with $\alpha$ running over the one-quark one-gluon and one-quark two-gluon components, and $u$ the parameter previously introduced in  Eq.~(\ref{def:CF1}). Quark flavor is fixed since the Hamiltonian 
cannot change it. 

The wf component without gluons is therefore solely specified  by the momentum and spin of the quark. We prepare each degree of freedom separately. The momentum basis states,
$\big|m^{1q}_1\big\rangle = \frac{1}{\sqrt{2}} \left(\ket{1}+\ket{-1}\right)$, $\big|m^{1q}_2\big\rangle = \frac{1}{\sqrt{2}}\left(\ket{2}+\ket{-2}\right)$, are thus combined with the two spin/polarization states and a fixed color state to give a total of 4 amplitudes.

For the one-gluon component, the gluon and quark momentum-basis states directly correspond to the values listed in Table~\ref{tab:discretization}. This results in four amplitudes for the gluon and four amplitudes for the quark. Their colors are combined into a triplet representation, and the state is fixed so as to have the same total color as the one-quark component. Including the five spin amplitudes associated with the gluon and quark leads to 13 amplitudes.

The wavefunction of the one-quark two-gluon sector must be symmetric under gluon exchange. The momentum basis states for the two gluons correspond to the ten symmetric combinations of the values listed in Table~\ref{tab:discretization}. Regarding color, combining the gluon octets together with the color triplet of the quark generates three independent triplet representations, yielding three independent states with the same overall color as the other sectors. These must be combined with the six symmetric spin states and three antisymmetric spin states in such a way that the total wavefunction is symmetric under simultaneous exchange of gluon color and spin/polarization degrees of freedom, see Table \ref{tab:1q2g-symmetry}. Including the four momentum basis states and the two spin states of the quark gives a total of 28 amplitudes.

\begin{table}[]
\caption{Symmetry under gluon exchange and number of amplitudes for the one-quark two-gluon sector. Spin and color are combined so that the overall wavefunction is symmetric under exchange.}
\label{tab:1q2g-symmetry}
\begin{tabular}{c||ccc}
$1q2g$ Sector & Momenta & Polarization & Color \\ 
\hhline{=#===}
\begin{tabular}[c]{@{}c@{}}Sector\\ Symmetry\end{tabular} & S & S/A & S/A \\ 
\hhline{-||---}
\begin{tabular}[c]{@{}c@{}}Combined \\ Symmetry\end{tabular} & S & \multicolumn{2}{c}{S} \\ 
\hhline{-||---}
Amplitudes & 10 & \multicolumn{2}{c}{12}
\end{tabular}
\end{table}

Altogether, the variational wavefunction with a variable number of gluons contains $46$ real and $2$ complex amplitudes parameterized in terms of $37$ real variational parameters through trigonometric functions enforcing normalization.
\subsection{Hamiltonian decomposition}
\label{Hamiltonian-decomposition}
The Hamiltonian terms appearing after the expansion in creation and annihilation operators (see appendix) need to be encoded into transformations among qubits before applying any algorithm for the measurement of expectation values.  Encoding allows for the decomposition of operators in terms of sums of tensor products of Pauli matrices, known as Pauli strings. A straightforward measurement algorithm, and the one we use here, consists of sampling these Pauli strings and reconstructing the expectation values by combining the individual results. In the following, we illustrate several simplifications that can be applied to the decompositions using representative terms of the Hamiltonian and leave a more detailed discussion about the scaling and feasibility of this approach to section \ref{scaling}.

\subsubsection{Kinetic energy terms}
The simplest operator in the momentum basis is the kinetic energy
\begin{equation}
    H_{g11} = \int \frac{d^3p}{2E_p}\,h_{g11}(p;i,j)\,a^{\dagger}_{pia}a_{pja}\ ,
\end{equation}
where $h_{g11}(p;i,j)$ can be found in the supplementary material along with an explanation of the subindex notation. Note that, as a consequence of the gluon polarization basis Eq.~(\ref{def:gluon-pol}), which differs from the usual decomposition in transverse and orthogonal polarizations, the incoming and outgoing gluon polarization states can differ and therefore the term is not diagonal.

Its expectation value is proportional to the matrix element of the operators over the variational wavefunctions
\begin{equation}
    \hat{H}_{g11} \propto \left\langle \Psi\right|a^{\dagger}_{pia} a_{pja} \left|\Psi\right\rangle\  = \langle a^{\dagger}_{pia} a_{pja} \rangle,
\end{equation}
which, applying the PRE, reads
\begin{align}
\langle a^{\dagger}_{pia} a_{pja} \rangle =& \sum^{n_g}_{l} \langle\stepS{l} \cdot \projector{n_g-l}{0}\nonumber \\
&\otimes(\ketbrac{1}{1}\otimes \set{pia}\scrap{pja})_l\otimes\projector{l-1}{l-1}\cdot \stepS{l}\rangle\ ,
\end{align}
where the factors associated with the creation and annihilation operators combine into $(\ketbrac{1}{1}\otimes \set{pia}\scrap{pja})_l$. Several simplifications follow immediately: because the PRE stores particles in contiguous registers, the projectors acting on registers to the right of the $l$-th register automatically evaluate to one, and can be replaced by the identity. The same argument reduces the projectors on the left to a single projector. Furthermore, since the wavefunction is symmetric under particle exchange, the step symmetrizers $\stepS{l}$ just reduce to a factor of $\sqrt{l}$ each,
\begin{equation}
\langle a^{\dagger}_{pia} a_{pjb} \rangle = \sum^{n_g}_{l} l\,\langle \projector{1}{0}\otimes(\ketbrac{1}{1}\otimes \set{pia}\scrap{pja})_l\rangle\ .
\label{eq:adga-PRE}
\end{equation}
Its interpretation is now straightforward: the operator checks whether register $l$ is occupied and register $l+1$ is empty. This implies that there are exactly $l$ particles by the filling order of the PRE and, as a consequence, the measured energy has to be multiplied by $l$.  Owing to the symmetrization of the wavefunction, this is equivalent to measuring the energy on every register. 

Finally, we use the encoding of Table \ref{tab:discretization} to decompose Eq.~(\ref{eq:adga-PRE}) into Pauli strings. The decomposition of projectors and single-qubits operators $C_{ij}$ is straightforward as per definitions Eq.~(\ref{def:Cij}) and Eq.~(\ref{eq:projector-explicit}), so we concentrate on the sum over quantum-number values over the $s^{\dagger}s$ product, taken into account the encodings of tables \ref{tab:encoding} and \ref{tab:discretization},
\begin{align}
\sum_{pija}& h_{g11}(p;i,j)\,\set{pia}\scrap{pja} \nonumber\\
= &\,\sum_{pij}\,h_{g11}(p;i,j)\,\left\{\set{p}\scrap{p}\otimes\set{i}\set{j}\right\}\otimes\sum_a\set{a}\scrap{a} \nonumber \\
= &\,\left\{\frac{1}{4}(II+ZI)\otimes\left[I\,h^{+}_{g11}(1)+Z\,h^{-}_{g11}(1)\right.\right. \nonumber\\  
& \left. + X\,h_{g11}(1,1,2)\right]\nonumber \\
&+\frac{1}{4}(II-ZI)\otimes\left[I\,h^{+}_{g11}(2)+Z\,h^{-}_{g11}(2)\right.\nonumber\\ 
& \left. + X\,h_{g11}(2,1,2)\right]\bigg\}\otimes III\ ,
\end{align}
where $h^{\pm}_{g11}(p) = h_{g11}(p,1,1) \pm h_{g11}(p,2,2)$ and we took into account that $h_{g11}(p,i,j) = h_{g11}(-p,i,j)$ and $h_{g11}(p,i,j) = h_{g11}(p,j,i)$. Since momenta are only along one direction, the terms $h_{g11}(p,i,j)$ with $i\neq j$ vanish and the number of Pauli strings reduces to 4. Adding the contributions from projectors and $C_{ij}$, the total Pauli count is $16\,n_g$.

\subsubsection{Anomalous gluon-pair term}
The Hamiltonian operator and squared Gauss' law constraint have   anomalous terms that create and annihilate particles from the vacuum, for example
\begin{equation}
    H_{g20} = \int \frac{d^3 p}{4 E_p} \, h_{g20}(p;i,j)\,  (a^{\dagger}_{pia}a^{\dagger}_{-pja} + \text{h.c.})\ .
\end{equation}
Such operators do not conserve particle number and their eigenstates are therefore superpositions of states containing arbitrarily many gluons. As a consequence, they cannot be represented exactly within a truncated Fock space with a bare vacuum, and one must resort to nonperturbative techniques such as the Bogoliubov-Valatin transformation. This will be considered in future work.

Introducing the PRE expansion for the creation and annihilation operators and simplifying the step-symmetrizers and projectors, the expectation value reads
\begin{align}
\langle &a^{\dagger}_{pia} a^{\dagger}_{-pja} + \text{h.c.}\rangle =  \sum^{n_g}_{l}\sqrt{l\,(l-1)}\,\langle \projector{1}{0}\otimes \nonumber \\
&\left[(\ketbrac{1}{0}\otimes \set{pia})_l\otimes (\ketbrac{1}{0}\otimes \set{-pja})_{l-1}+\text{h.c.}\right]\otimes\projector{1}{1}\rangle\ .
\end{align}
The main differences are that being a two-body operator, it changes the state of two registers at once, as seen by the tensor product $(...)_l\otimes(...)_{l-1}$. The set operators and control qubits associated with the creation operators activate two presence qubits and initialize the quantum numbers of two previously empty registers. In contrast, the Hermitian conjugate removes two particles and restores all register qubits to state $\ket{0}$.

Before the Pauli expansion, we group the operators within the two registers depending on the quantum number in which they act,
\begin{align}
\sum_{pija}&h_{g20}(p,i,j)\,(C_{10}\otimes\set{pia})_{l+1}\otimes(C_{10}\otimes\set{-pja})_l+\text{h.c.}\nonumber\\
& = \bigg\{C_{10}\otimes C_{10}\bigg\}_{\mathcal{P}}\bigg\{\sum_{pij}h_{g20}(p,i,j)\,\set{pi}\otimes\set{pj}\bigg\}_\text{ms}\nonumber\\
&\bigg\{\sum_a\set{a}\otimes\set{a}\bigg\}_\text{c} + \text{h.c.}\,,
\label{eq:h20-expansion}
\end{align}
where, as in Eq.~(\ref{def:single-particle-set}), $c$ stands for color, $ms$ for momentum-spin and $\mathcal{P}$ for presence. Within each curly-brace grouping, the factors on the left (right) of the tensor product act on register $l+1$ ($l$), respectively. To ease notation, from now on we omit the subindices in the braces. 

With the discretization of  Table \ref{tab:discretization}, a naïve expansion of the operator into Pauli strings gives $\mathcal{O}(10^6)$ terms, while the color set operators produce $512$ Pauli strings on their own. To reduce the cost, we can take advantage of the definite state of empty registers. Consider the following expectation value 
\begin{equation}
\langle C_{10}\otimes\ket{1}\bra{0} + \text{h.c.}\rangle = \langle \frac{X-iY}{2}\otimes \frac{X-iY}{2} + \text{h.c.}\rangle,
\end{equation}
which can appear, for example, because of the spin set operators, and will decompose into four Pauli strings. We can reduce the number of strings noticing that, by construction, mode qubits of empty registers are always in state $\ket{0}$, so the expectation values 
\begin{equation}
\langle C_{10}\otimes\ket{0}\bra{1} + \text{h.c.}\rangle,\,\,\langle C_{10}\otimes\ket{1}\bra{1} + \text{h.c.}\rangle,
\end{equation}
will always yield zero. Remarkably, the combined operator
\begin{equation}
\langle C_{10}\otimes(\ket{0}\bra{1} + \ket{1}\bra{0})+  \text{h.c.}\rangle = \langle C_{10}\otimes X +  \text{h.c.}\rangle ,
\label{eq:pauli-reduction}
\end{equation}
requires fewer Pauli strings. Such type of redundant terms can be added to $\textit{set}$ and $\textit{scrap}$ operators among multiple mode qubits, such as the color and momentum-spin parts of Eq.~(\ref{eq:h20-expansion}). Doing so drastically reduces the number of Pauli strings to $\mathcal{O}(10^2)$. Note, however, that for Eq.~(\ref{eq:pauli-reduction}) to remain true, the wavefunction must be prepared ensuring that while the presence qubit is in state $\ket{0}$, every other mode qubit is also in state $\ket{0}$. Any leakage at the preparation step will introduce spurious contributions to the expectation value, which would otherwise be avoided with the complete Pauli string expansion.
\begin{widetext}
\subsubsection{Gluon-pair interaction}
The last term we consider is (we omit the Dirac delta function of momentum conservation)
\begin{equation}
H_{g22}=\frac{g^2}{2}\int\left[pqkl\right]\bigg( f^{abc}f^{ade}\delta_{i'}^{i}\delta_{j'}^{j}+f^{abe}f^{adc}\left(\delta_{i'}^{i}\delta_{j'}^{j}-\delta_{j}^{i}\delta_{i'}^{j'}\right)\bigg) a_{pib}^{\dagger}a_{kjc}^{\dagger}a_{qi'd}a_{lj'e},
\end{equation}
where $[pqkl]$ is defined in the appendices. This interaction term also involves two registers, but in contrast to the previous case, they both remain occupied. 

After applying the PRE and simplifying projectors and step-symmetrizers the result reads
\begin{equation}
\langle a_{pib}^{\dagger}a_{kjc}^{\dagger}a_{qi'd}a_{lj'e}\rangle = \sum_l \left(l+1\right)\,l\,\langle\projector{1}{0}\otimes\left(\ketbrac{1}{1}\otimes\set{pib}\scrap{lj'e}\right)_{l+1}\otimes\left(\innerid\otimes\set{kjc}\scrap{qi'd}\right)_{l}\rangle.
\end{equation}
To continue, we separate the expectation values into two terms and reorganize the set and scrap operators as before
\begin{align}
\langle H_{g22} \rangle_1  &= \sum_ll(l+1)\big\langle\projector{1}{0}\otimes\big\{C_{11}\otimes\mathfrak{i}\big\}\bigg\{\frac{g^2}{2}\sum \,\delta^{i}_{i'}\,\delta^{j}_{j'}\,\set{pi}\scrap{lj}\otimes\set{kj}\scrap{qi'}\bigg\}\big\{\sum f^{abc}f^{ade}\set{b}\scrap{e}\otimes\set{c}\scrap{d}\big\}\big\rangle,\\
\langle H_{g22}\rangle_2& = \sum_ll(l+1)\big\langle\projector{1}{0}\otimes\big\{C_{11}\otimes\mathfrak{i}\big\}\bigg\{\frac{g^2}{2}\sum \,(\delta^{i}_{i'}\,\delta^{j}_{j'}-\delta^{i}_{j}\,\delta^{i'}_{j'})\set{pi}\scrap{lj}\otimes\set{kj}\scrap{qi'}\bigg\}\big\{\sum f^{abe}f^{adc}\set{b}\scrap{d}\otimes\set{c}\scrap{d}\big\}\big\rangle,
\end{align}
\end{widetext}
The fact that the color and momentum-spin parts are factorized allows for a partial diagonalization of the interaction. We first write the different combinations of color states over two registers into states over a single composite register
\begin{align}
C_1(r,r')\,\set{r}\scrap{r'} & = \sum_a  f^{abc}f^{ade}\set{b}\scrap{e}\otimes\set{c}\scrap{d},\label{eq:diag1} \\
C_2(r,r')\,\set{r}\scrap{r'} &= \sum_a  f^{abe}f^{adc}\set{b}\scrap{e}\otimes\set{c}\scrap{d},\label{eq:diag2}
\end{align}
where the color indices $b,c,d,e$ are combined into shorthand indices $r,r'$  related by the function 
\begin{equation}
    f(x,y) =(N_c^2-1)(x-1)+y-1\ ,
\end{equation} 
so that $r=f(b,c)$ and $r' = f(e,d)$. 
The matrices $C_1$ and $C_2$ are determined by the SU(3) color algebra, in particular, the structure constants and the encoding of color states. Therefore, they can be diagonalized once and for all as a classical preprocessing step. The unitary transformation that performs the diagonalization, $U_1$ and $U_2$, act only on the six color qubits associated with pairs of gluons and are independent of the momentum discretization and variational parameters.

Although implementing a generic unitary transformation on $n$ qubits requires $\mathcal{O}(n^2 4^n)$ elementary gates, this estimate does not introduce an exponential overhead here. The number of qubits in which $U_1$ and $U_2$ act remains fixed, while the number of applications scales only with the number of gluon pairs $\mathcal{O}(n_g^2)$.

The final expectation values are
\thinmuskip=0mu
\begin{align}
\langle H_{g22} \rangle_{1(2)} = \sum_l &l(l+1)\big\langle\projector{1}{0}\otimes\big\{\cdots\big\}\big\{\cdots\big\}\nonumber\\
&\big\{ U_{1(2)}^{\dagger}\,\big(\sum_{r} D_{1(2)}(r)\,\set{r}\scrap{r}\big)\,U_{1(2)}\big\}\big\rangle,   
\end{align}
\thinmuskip=3mu
where the matrices $D_{1}$ and $D_{2}$ arise from the diagonalization of $C_{1}$ and $C_{2}$, respectively.

This transformation leads to a substantial reduction in the Pauli string expansion. Prior to diagonalization, Eqs.~(\ref{eq:diag1}, \ref{eq:diag2}) contain $1349$ and $590$ terms, respectively, while after diagonalization these are reduced to $8$ and $62$ terms.

\section{Numerical simulations}
\subsection{Simulation Parameters and Subtraction}
We wish to set the computer simulation parameters to a regime where quark matter can have set in, so that we do not have to worry too much about clustering to form baryons at this stage~\cite{baym_hadrons_2018},
but that is not too far from densities expected in neutron star cores (whether~\cite{Ayriyan:2025rub} or not~\cite{Brandes:2023bob} the quark matter regime is reached in a star is irrelevant from the point of view of constraining the EoS from theory alone; and perhaps high-enough densities could be approached in dynamical settings such as rotating stars~\cite{Moreno:2023xez,Gartlein:2025sko,Bagchi:2026rwj} or mergers~\cite{Chatterjee:2025zrh} anyway). We denote this set by the moniker  ``core''.

Due to the limited amount of simulatable qubits, the particle number $N$ remains fixed. Consequently, the system density $n = N/V$ is uniquely determined by the volume, which is in turn governed by the momentum cutoff $\Lambda$, see Eq.~(\ref{eq:volume}). To ensure our thermodynamic consistency, the chemical potential $\mu_B$ cannot be chosen arbitrarily and must be tuned as a dependent function of $\Lambda$. In a degenerate Fermi gas, density scales as $n \propto \mu_B^3$. Since the volume scales as $V \propto \Lambda^{-3}$, the chemical potential is taken to be a linear function of $\Lambda$:
\begin{equation}
\mu_B(\Lambda) = \mu_{B,\text{core}} + \frac{\mu_{\text{match}} - \mu_{\text{core}}}{\Lambda_{\text{match}} - \Lambda_{\text{core}}} (\Lambda - \Lambda_{\text{core}}) \, .
\end{equation}

The renormalization scale is set to be twice the quark chemical potential $\mu_R = 2 \mu_B/3$, so the coupling constant also becomes a function of $\Lambda$. We use the one-loop perturbative formula with $\Lambda_{\text{QCD}} = 0.25 \text{ GeV}$ to calculate $\alpha_S(\mu_R(\Lambda))$. Table \ref{tab:parameters} shows typical values of the baryon chemical potential and $\alpha_S$ for different densities.

\begin{table}[]
\caption{Number-density values in units of the nuclear saturation density $n_s = 0.15\,\text{nucleons/fm}^3$ and baryon chemical potential corresponding to the onset of perturbation theory in quark matter, and the limiting calculation of \cite{gorda_soft_2021}. The coupling constant is calculated at one loop taking $\Lambda_{QCD} = 0.25\,\text{GeV}$ and a renormalization scale $\mu = 2\mu_B/3$.}
\label{tab:parameters}
\begin{tabular}{c||c|c|c}
Sets & $n/n_s$ & $\mu_B$ [GeV]& $\alpha_S$ \\ \hhline{=#= = =}
Core  & 7   & 1.4             & 0.53       \\
\hhline{-||- - -}
Middle & 20 & 2 & 0.41 \\
\hhline{-||- - -}
Match & 42     & 2.6             & 0.36       \\
\end{tabular}
\end{table}

Although we have not attempted a full renormalization of the QCD Hamiltonian, a divergent constant term requires a counterterm, which is probably required in any conceivable renormalization; such constants are set {\it in vacuo}, independently of the medium parameters.
To set the correct energy and pressure at one point,  the matching scale to pQCD, we apply the following subtraction to the Hamiltonian
\begin{eqnarray}
H\rightarrow H' = H-H_0-(h_0+h_2\,g(\Lambda)^2)\,\Lambda\ ,
\label{def:renorm}    
\end{eqnarray}
where $H_0$, $h_0$, and $h_2$ are constants fixed through high-density perturbative QCD \cite{gorda_soft_2021} results.

To estimate the derivative of $H$ as a function of $\Lambda$ at the matching scale we fit the minimized energy (parametrized by coefficients which we delay to table \ref{tab:dmatch}) to the function 
\begin{equation}
    f_1(\Lambda;a_1,b_1) = b_1\, E_0\, (\Lambda/\Lambda_0)^{a_1},
    \label{eq:fit1}
\end{equation}
where $E_{\text{min}}$ and $\Lambda_{\text{min}}$ are the minimum values of the energy and momentum scales been fitted, they are included to make the fitting parameters dimensionless. The derivative are then analytically calculated and we solve for the constants $H_0$, $h_0$ and $h_2$ of Eq.~(\ref{def:renorm}) so as to reproduce a pressure of $p=4200\,\text{MeV/fm}^3$ at an energy density $\epsilon = 12600\, \text{MeV/fm}^3$, the matching scale to~\cite{gorda_soft_2021}, while simultaneously setting a vanishing expectation value over the perturbative vacuum for the free Hamiltonian, 
\begin{equation}
\left\langle v\right|H'(g=0,\Lambda=\Lambda_{\text{match}})\left|v\right\rangle\  = 0.
\end{equation} 

\begin{figure}[h!]
\centering
\includegraphics[width=0.45\textwidth]{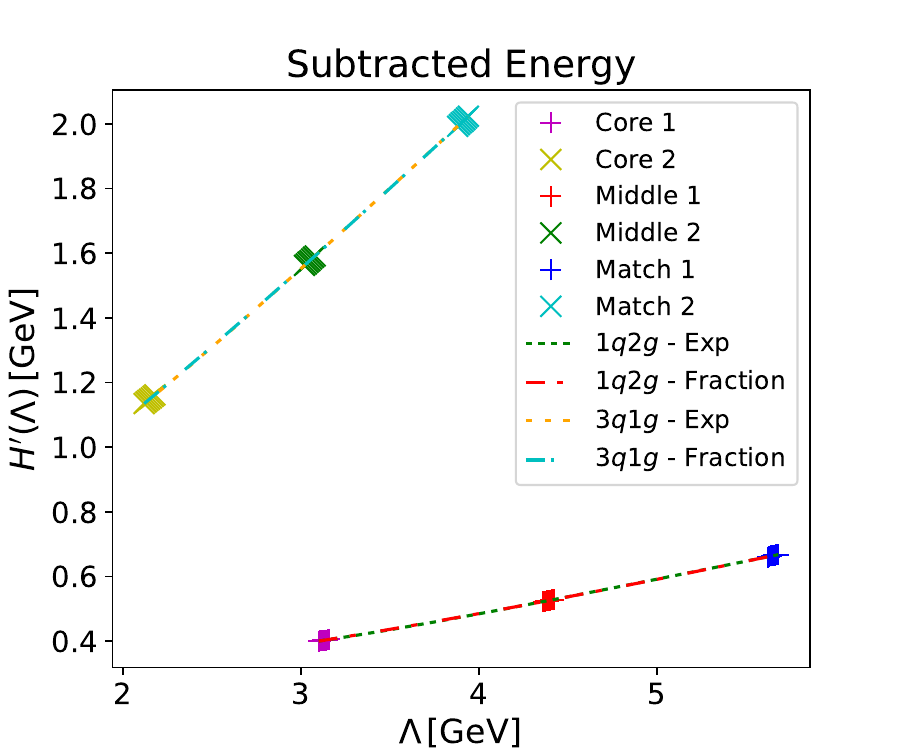}
\caption{Energies obtained after applying Eq.~(\ref{def:renorm}) to the minimum of the expectation value of the Hamiltonian constrained by Gauss' law. We see no sensitivity to changing the fitting procedure, ``Exp'' denoting an exponential function fit, see Eq.~(\ref{eq:Exp-fit}) and ``Fraction'' a fit with a fraction of polynomials, see Eq.~(\ref{eq:Fraction-fit}). We do see strong sensitivity to the trial variational wavefunction. This is expected, and a working large-scale quantum computer, with its massive Hilbert space, would be the tool to address it.}
\label{fig:Subtracted-energy}
\end{figure}

The resulting coefficients for the wavefunctions described in subsection~\ref{sec:varwaves} are listed in table \ref{tab:subtraction-params}. Eq.~(\ref{def:renorm}) then gives the subtracted energies $H'$ seen in Fig.~(\ref{fig:Subtracted-energy}),  from which we calculated the Equation of State and the sound velocity squared, $c_s^2$.

\begin{table}[]
\caption{Subtraction constants defined in Eq.~(\ref{def:renorm}) and fit so the energy density matches that computed with pQCD methods~\cite{gorda_soft_2021} at a high scale, with $h_0$ and $h_2$  dimensionless. }
\label{tab:subtraction-params}
\begin{tabular}{c||c|c|c}
Trial wf & $H_0\,[GeV]$ & $h_0$ & $h_2$ \\ \hhline{=#= = =}
$3q1g$  & 0.84   & 11.79            & 1.55      \\
\hhline{-||- - -}
$1q2g$ & 1.33 & 11.76 & 1.59 \\
\end{tabular}
\end{table}

\subsection{Results}
The EoS and sound velocity squared can be obtained from the derivatives of the fit to the subtracted energy data, if such fits are denoted by $H'(\Lambda)$, from Eq.~(\ref{eq:volume}) and Eq.~(\ref{def:pres-energydensity}) we have,
\begin{equation}
    p = \frac{1}{3}\frac{\Lambda}{V}\frac{\partial H'}{\partial \Lambda},
\end{equation}
and 
\begin{equation}
    c^2_s = \frac{\Lambda}{3}\frac{\Lambda\,\partial^2 H'/\partial \Lambda^2+4\,\partial H'/\partial \Lambda}{\Lambda\, \partial H'/\partial \Lambda+3\,H'}\, .
    \label{eq:finalcs}
\end{equation}

\begin{figure}[h!]
\centering
\includegraphics[width=0.43\textwidth]{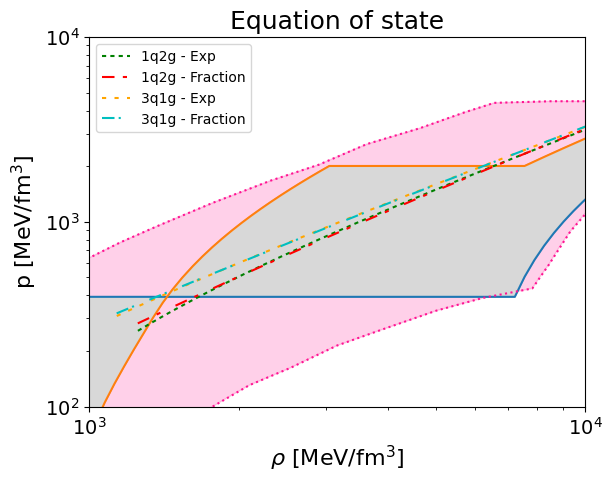}
\includegraphics[width=0.45\textwidth]{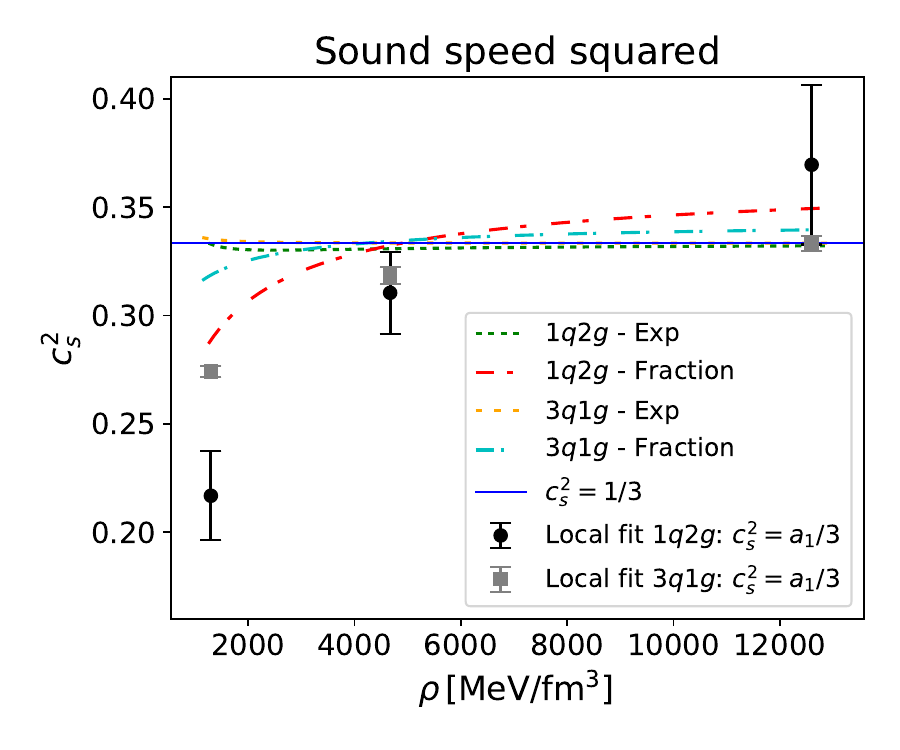}
\caption{Pressure and sound velocity squared as functions of the energy density, obtained from the minimum of the time-axial QCD Hamiltonian expectation value evaluated with the variational wavefunction described in Sec.~\ref{vwave:3q1g} and ~\ref{vwave:1q2g}, and after applying Eq.~(\ref{def:renorm}). }
\label{fig:EoS1}
\end{figure}

To assess the uncertainty in the fitting procedure we considered two different functions,
\begin{equation}
    f_2(\Lambda;a_2,b_2,c_2) =a_2\,E_0\, (\Lambda/\Lambda_0)\,e^{b_2\,(\Lambda/\Lambda_0)^{-c_2}},
    \label{eq:Exp-fit}
\end{equation}
\begin{equation}
    f_3(\Lambda;a_3,b_3,c_3) = a_3\,E_0\,\frac{1+b_3\,(\Lambda/\Lambda_0)^2}{1+c_3\,(\Lambda/\Lambda_0)},
    \label{eq:Fraction-fit}
\end{equation}
denoted by ``Exp'' and ``Fraction'' respectively. The constants $E_0$ and $\Lambda_0$ are the minimum values of the energy and momentum scale being fitted and make the fitting parameters dimensionless. They are both fitted to the subtracted data of Tables \ref{tab:dcore}, \ref{tab:dmid} and \ref{tab:dmatch} in appendix \ref{mindata}. The fitting parameters for both wavefunctions are listed in Tables \ref{tab:fitting-1} and \ref{tab:fitting-2}. The results for the EoS and velocity of sound can be found in Fig.~(\ref{fig:EoS1}).

\begin{table}[]
\caption{Fitting parameters of the function Eq.~(\ref{eq:Exp-fit}) for both wavefunctions. Errors correspond to the square roots of the diagonal elements of the covariance matrix. All parameters are dimensionless.}
\label{tab:fitting-1}
\begin{tabular}{c||c|c|c}
Trial wf & $a_2$ & $b_2$ & $c_2$ \\ \hhline{=#= = =}
$3q1g$  & $0.9552(2)$   & $0.0460(2)$           & $3.98(5)$     \\
\hhline{-||- - -}
$1q2g$ & $0.900(1)$ & $0.106(1)$ & $3.6(1)$ \\
\end{tabular}
\end{table}

\begin{table}[]
\caption{Fitting parameters of the function Eq.~(\ref{eq:Fraction-fit}) for both wavefunctions. Errors correspond to the square roots of the diagonal elements of the covariance matrix. All parameters are dimensionless.}
\label{tab:fitting-2}
\begin{tabular}{c||c|c|c}
Trial wf & $a_3$ & $b_3$ & $c_3$ \\ \hhline{=#= = =}
$3q1g$  & $0.976(7)$ & $5.29(8)$ & $5.14(12)$     \\
\hhline{-||- - -}
$1q2g$ & $0.918(14)$ & $1.80(8)$ & $1.57(11)$  \\
\end{tabular}
\end{table}

The velocity of sound is also estimated from local fits. The three sets of points ``core'', ``middle'', and ``match'' were separately fitted with Eq.~(\ref{eq:fit1}).  The velocity of sound, Eq.~(\ref{eq:finalcs}) is, for this specific case, one third of the exponent, $c_s^2= a_1/3$. The fitting parameters are listed in Table \ref{tab:local-fits} and the corresponding sound velocities squared plotted in Fig.~\ref{fig:EoS1} as gray squares and black dots.

\begin{table}[]
\caption{Fitting parameters of function Eq.~(\ref{eq:fit1}) to the three data sets ``core'', ``middle'' and ``match''. All parameters are dimensionless and errors correspond to the square roots of the diagonal elements of the covariance matrix.}
\label{tab:local-fits}
\begin{tabular}{c|c||c|c}
Trial wf & Data set & $a_1$ & $b_1$ \\ \hhline{= =#= =}
 & Core   & $0.822(8)$ & $1.00019(13)$ \\
$3q1g$   & Middle   & $0.955(12)$ & $0.981(4)$ \\
& Match  & $1.00(1)$& $0.959(6)$   \\
\hhline{- -||- - }
 & Core   & $0.65(6)$ & $1.0013(7)$  \\
$1q2g$   & Middle   & $0.93(6)$ & $0.95(2)$ \\
& Match  & $1.11(11)$ & $0.85(6)$   \\
\end{tabular}
\end{table}

The results appear reasonable in the sense that they are compatible with the causality and thermodynamic consistency constraints discussed in \cite{Alarcon:2024hlj}. Moreover, the sound velocity is broadly compatible with its expected increasing behaviour towards the conformal gas (free QCD should have $c_s^2=1/3$, and scale-dependent vs. scale-invariant properties of neutron star matter are of great interest~\cite{Ivanytskyi:2024zip,Ma:2026glg,Xiong:2025jxq}).
While the uncertainty induced on the energy density by the fit (\emph{Exp} or \emph{Fraction}) is quite small at the level of the EoS, it becomes very apparent when taking a derivative of the pressure to yield the sound speed. Still, the calculation can only be considered a demonstration of the method because of the strong sensitivity to the variational wavefunction. Quantum computing is indeed the tool to address this in the future because of the vast spaces span by $O(10^4)$ qubits.

Further remarks concerning the systematic uncertainties of this exploratory calculation follow: firstly, the discretization introduced in Table~\ref{tab:discretization} should be extended to include multidimensional momentum configurations. Secondly, a quantum memory with capacity for configurations containing at least a few nucleons would be necessary to capture collective many-body effects. Third, the consequences of the Fock-space truncation are not yet fully understood. In particular, it is unclear whether enlarging the variational basis to include additional particles would allow the gauge to be fixed completely, or equivalently, whether the residual violation of Gauss' law observed in the present calculation is merely an artifact of the truncation. A possible alternative would be to formulate the problem directly in Coulomb gauge, where the constraint can be solved explicitly at the Hamiltonian level~\cite{lee_particle_1988}. Lastly, the naïve subtraction introduced in Eq.~(\ref{def:renorm}) should be replaced by a proper renormalization procedure~\cite{Robertson:1998va}.

\section{Outlook: Computational costs and necessary hardware}
\label{scaling}
The practical usefulness of the proposed encoding ultimately depends on how the computational resources scale. We therefore analyze the asymptotic growth of both the Hamiltonian and the squared Gauss operator, identify the interactions responsible for the dominant computational cost, and investigate whether physically motivated approximations can reduce that scaling. We count the number of non-vanishing matrix elements and the number of Pauli strings associated to each term of the Hamiltonian $H$ and (squared) Gauss operator $\mathcal{G}^2$, varying the size of the momentum lattice (characterized by $N$ the number of valid discretized momenta per dimension). We increase $N$ up to $N=32$ in one dimension.

In the left panels of Figs.~\ref{fig:Matrix elements scaling}~and~\ref{fig:Pauli strings scaling}, it can be ascertained that the growth with the momentum grid of both the number of matrix elements and that of Pauli strings is considerable, but still polynomial $\mathcal{O}(N^k)$. However, instead of considering the raw operators, we propose a truncation method on the matrix elements so that we eliminate the least relevant contributions.

We define a cutoff $\varepsilon$ equal to a fraction $\eta =10\%$ of the average of the absolute value of the momentum-space interaction-vertex coefficient $|c_\text{mom}|$ (consider the part depending on kinematics but preserve independence on the color algebra). The mathematical definition for $\varepsilon$ is
\begin{equation}
    \varepsilon = \eta \overline{|c_\text{mom}|}.
    \label{momentum cutoff}
\end{equation}

Note that the values taken by $\varepsilon$ vary with the momentum grid size $N$ and are indicated in Tab.~\ref{tab:epsilon_values}.

\begin{table}[h!]
\centering
\caption{Values of the cutoff $\varepsilon$ for different sizes of momentum lattice (given by $N$), as defined in Eq.~\eqref{momentum cutoff} and using $\eta = 0.1$.}
\label{tab:epsilon_values}
\begin{tabular}{c||c}
\hline
Grid size $N$ & Cutoff $\varepsilon$ [$\mathrm{GeV}$] \\ \hhline{=||=}
$2$ & $0.0475$ \\ \hline
$4$ & $0.0493$ \\ \hline
$8$ & $0.0648$ \\ \hline
$16$ & $0.0870$ \\ \hline
$32$ & $0.1167$ \\ \hline
\end{tabular}
\end{table}

By discarding the matrix elements satisfying $|c_\text{mom}|<\varepsilon$ before the construction of the Pauli string (a costly operation) we eliminate the terms suppressed by large momentum denominators (when $N$ is large) and preserve the algebra of the color structure.

This truncation does not act in the same way for every term:  those which have a strong contribution of energy --such as $q_{11}$ and $g_{11}$ in the Hamiltonian-- grow as momentum becomes larger, and their matrix elements satisfy $|c_\text{mom}|\geq \varepsilon$, so their scaling curves remain the same. They also make the mean of the $|c_\text{mom}|$'s larger with energy, which means that the relevant energy scale is higher and that terms must have a higher energy not to be neglected by $\varepsilon$. The interactions (i.e.,$g_{11}q_{11}$, $g_{20}q_{11}$ and $q_{22}$) are highly suppressed or even completely eliminated, as seen in the right panel of Fig.~\ref{fig:Matrix elements scaling}, whereas there is a significant reduction in the scaling of $g_{22}$ and $g_{21}$.

This has a direct effect of the scaling of the number of Pauli strings of the operator $H+G^2 (G=\mathcal{G}/\Lambda)$, which dictates the costs for the quantum computer. Table~\ref{tab:scaling-summary} shows an outline of the reduction in the power of the polynomial for the different interactions.

\begin{table}[t]
\centering
\caption{Polynomial scaling exponents $k$ for the number of matrix elements and Pauli strings of the combined operator $H+\mathcal{G}^2$ as a function of the momentum modes $N$, fitted as $\propto N^k$. Terms marked with ``--'' are completely eliminated by the cutoff.}
\label{tab:scaling-summary}
\begin{tabular}{c||c|c||c|c}
\hline
& \multicolumn{2}{c||}{Matrix elements} & \multicolumn{2}{c}{Pauli strings} \\
Term & $\varepsilon=0$ & $\varepsilon>0$ & $\varepsilon=0$ & $\varepsilon>0$ \\ \hhline{=#=|=||=|=}
$q_{11}$       & 1.0 & 1.0 & 0.6 & 0.6 \\ \hline
$g_{11}$       & 1.0 & 1.0 & 1.0 & 1.0 \\ \hline
$g_{10}q_{11}$ & 2.3 & 2.0 & 2.4 & 2.4 \\ \hline
$g_{11}q_{11}$ & 2.9 & -- & 3.3 & -- \\ \hline
$g_{20}$       & 1.0 & 1.0 & 1.0 & 1.0 \\ \hline
$g_{20}q_{11}$ & 3.4 & -- & 3.3 & -- \\ \hline
$g_{21}$       & 2.4 & 2.0 & 3.2 & 3.2 \\ \hline
$q_{22}$       & 2.5 & -- & 2.4 & -- \\ \hline
$g_{22}$       & 3.0 & 1.6 & 3.4 & 2.5 \\
\hline
\end{tabular}
\end{table}

With $\varepsilon =0$, the number of Pauli strings for $N=32$ is over $\sim 10^7$ mainly because of the $g_{20}g_{11}$ terms. The cutoff completely eliminates it and the number of Pauli string is reduced to $\mathcal{O}\sim 10^6$. This cut-off reduces the number of Pauli strings associated to some terms that scaled very fast. We control the scaling but at the same time, we keep the physics at low and high momenta  unaltered. Therefore, we provide an approximation for the operators and improve the computational costs. We define the survival factor as the number of Pauli strings that remain after the cutoff; and it is illustrated in Fig.~\ref{fig:survivial rate terms}. If a term is unaffected by the cutoff, its survival factor will be 100\% whereas if it is  completely eliminated, it will have a reduction factor of 0\%. 

The terms corresponding to the dominant interactions ($g_{11},\, q_{11}$ and $g_{10}q_{11}$) have a reduction favor of 100\%. On the other hand, there are terms describing more suppressed interactions which, for $N=32$, reduce drastically. Note that, for a small number of terms, the number of Pauli strings actually slightly increases. This occurs because after eliminating some matrix elements, certain Pauli strings no longer simplify, as cancellation patterns change. In such cases, the cutoff is not worth applying and we just leave it out.

Although (a) these cutoffs improve the prospective computational cost and (b) moderately reduce the number of required Pauli strings, certain terms ({\it e.g.}, $g_{21}$) retain their original scaling. This highlights the importance of pre-measurement simplification strategies for further reducing the overall computational cost.

It will be profitable, once such computations become routinely accessible on operational quantum computers, to develop more refined criteria for selecting the most relevant terms and to develop further simplification strategies while preserving the essential physics. This is, in fact, one of the main motivations behind the present research: only a complete end-to-end implementation makes it possible to identify computational bottlenecks and uncover simplifications, such as those discussed in Sec.~\ref{Hamiltonian-decomposition}, that would otherwise remain hidden.

\subsection*{Conclusions}

In this article we have shown a beginning-to-end demonstration of how an eventual quantum computer can be employed to extract the Equation of State of Neutron Star Matter relevant for neutron--star, directly from Quantum Chromodynamics, helping the field progress beyond the Nambu-Jona-Lasinio model~\cite{Buballa:2003qv} and effective theory in a regime where it may not be applicable.
We have done so employing the particle-register encoding previously developed at Complutense and standard methods for gauge-fixing and Hamiltonian encoding. While current hardware does not allow the necessary number of operations to be run with any reasonable fidelity
we have to content ourselves with classical-simulator runs addressing a unit cell with a few particles immersed in a chemical-potential bath at zero temperature. This will entail an advantage of the quantum-computer treatment over traditional lattice gauge theory calculations, which suffer from a sign problem. Although our setup is demonstrably running and allows the extraction of rather coarse equations of state, the limited memory of a classical computer cluster means that we need to wait for actual quantum computers in the thousands of qubits whose memory will span vast Hilbert spaces unassailable by currently existing methods, profiting from an undoubted quantum-memory advantage. Because there is a good chance that quantum computers at this scale will operate simultaneously with third-generation Gravitational Wave Observatories such as the Einstein Telescope~\cite{ET:2025xjr}  or others which have been proposed, we think this line of research should be further pursued.

\begin{acknowledgments}
We thankfully acknowledge the collaboration of Mr. \'Alvaro Gallardo Gallego in the initial phases of this project, particularly when exploring alternative quantum computing encodings (Section IV of the supplementary material) although most of his work has not been incorporated in the final project here reported. He is here credited in accordance with the Authorship Ethics Standard of the APS.
Work partially supported by grants 
PID2023-147072NB-I00; 
PID2022-137003NB-I00 
FPU21/04180 
and FPU24/00948 
of the Spanish MCIN/AEI/10.13039/501100011033/ and Ministry of Universities.
Numerical computing carried out at Granada's PROTEUS supercomputer.

\end{acknowledgments}

\onecolumngrid
    \begin{center}
        \nopagebreak
        \includegraphics[width=0.8\textwidth]{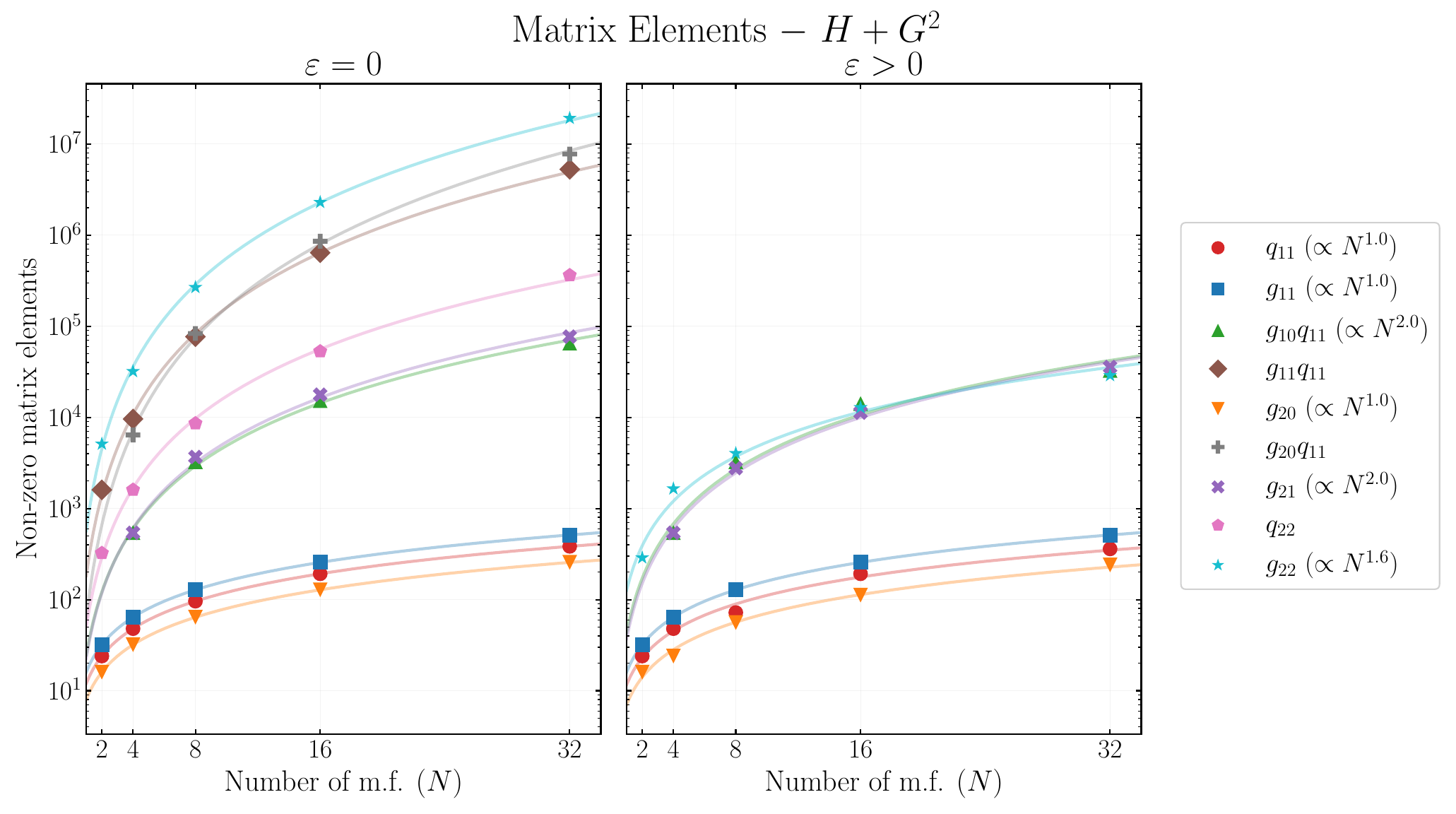}
        \vspace{8pt} 
        \captionof{figure}{Left: number of non-vanishing matrix elements for the combined operator $H+G^2$ associated to each interaction term against the number of momentum fractions (momentum grid size).  Right: remaining ones after applying a cutoff $\varepsilon$ given by~\eqref{momentum cutoff} setting those too small and compatible with noise to be exactly zero. The legend includes selected monomial fits for the scaling of  each interaction term with the number of modes after applying the cutoff $\varepsilon$.}
        \label{fig:Matrix elements scaling}
    \includegraphics[width=0.8\linewidth]{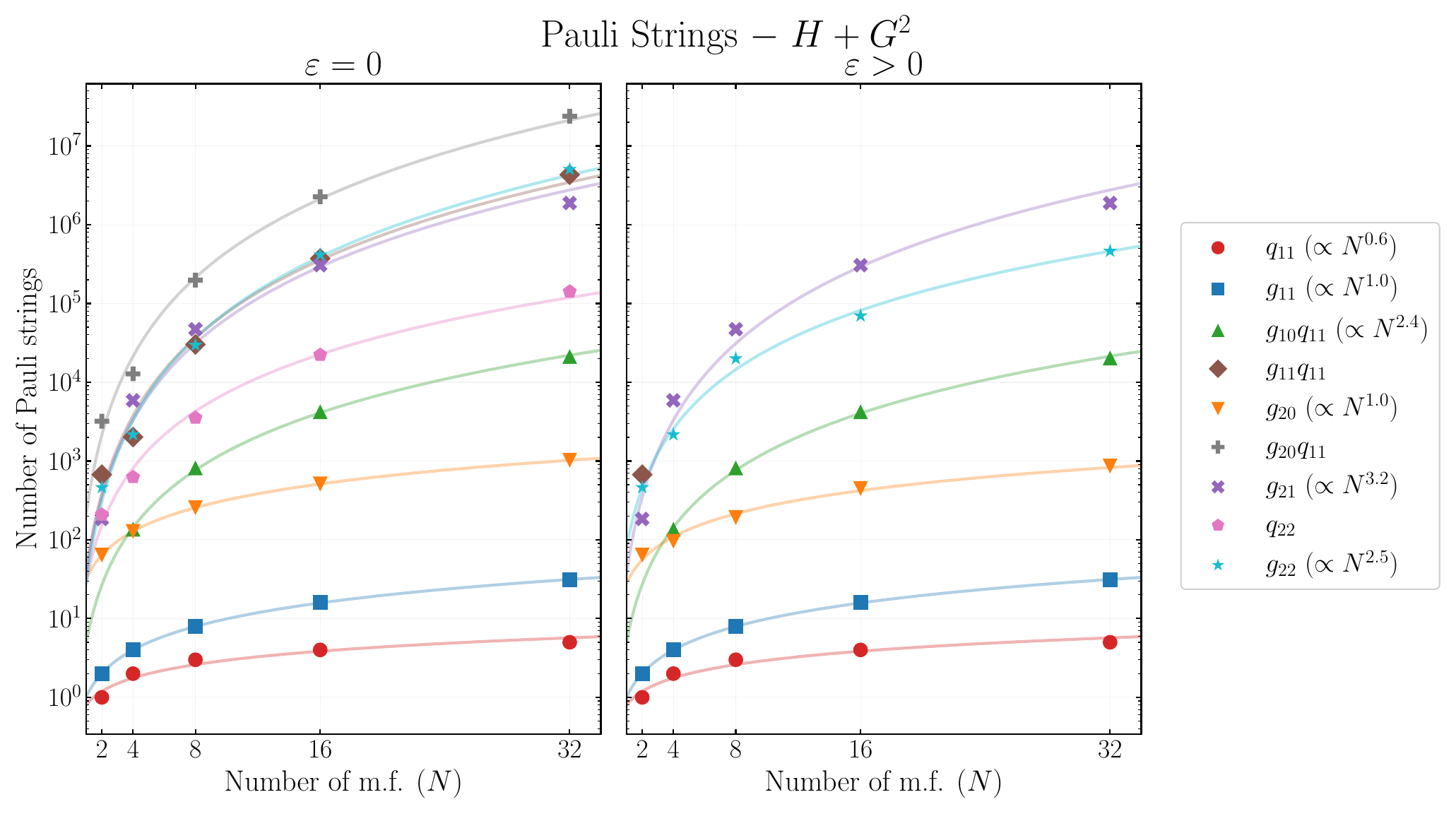}
    \captionof{figure}{Number of Pauli strings in the combined operator $H+G^2$, term by term (as function of the momentum fractions {\it i.e.} momentum grid size). Left: without applying any cutoff $\varepsilon$ (thus, the actual size of the computation). Right:  applying the cutoff given by~\eqref{momentum cutoff}.  Legend: Leading behavior of each interaction term's scaling after applying the cutoff $\varepsilon$.}
    \label{fig:Pauli strings scaling}. 
    \end{center}


\begin{center}
        \nopagebreak
    \includegraphics[width=0.8\linewidth]{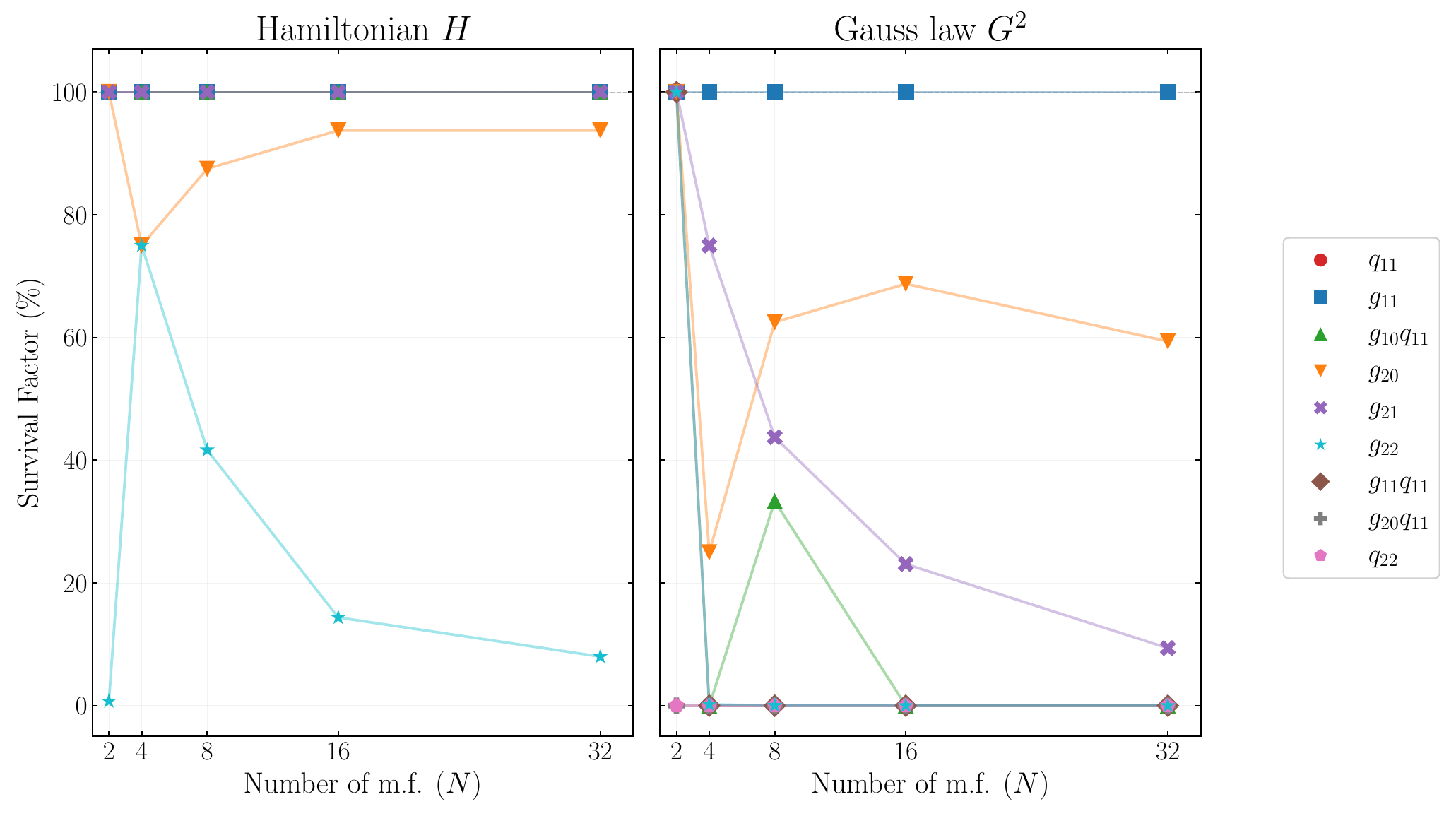}
    \captionof{figure}{Percentage of Pauli strings remaining after cutting off irrelevantly small matrix element,  as a funnction of the momentum fractions $N$ (the grid size).}
    \label{fig:survivial rate terms}
\end{center}
\twocolumngrid
\appendix
\onecolumngrid
\section{Complete expansion of the squared gauge-fixing operator}
\label{Apendix:Gauss}
In this section, we explicitly present the full expansion of the squared gauge-fixing operator $\mathcal{G}^2=\int d^3x\sum_a \mathcal{G}^a(x)\,\mathcal{G}^a(x)$  (the square not to be confused with the superindex $a$ labelling color) in Weyl's time-axial gauge, expressed in terms of the momentum-space particle creation and annihilation operators. As can be seen in Eq.~(7) of the main text, the mechanical dimensions of $\mathcal{G}^2$ are those of a squared color density multiplied by volume, thus $[E^{3}]$.

Before giving the full algebraic expansion --which contains a large number of interaction contributions-- we introduce a compact, systematic notation to classify the terms arising from $\mathcal{G}^2$ according to particle content and operator structure. These terms consist of products of creation and annihilation operators accompanied by their Hermitian conjugate (h.c), because of the self-adjointness of $\mathcal{G}^2$, and we pair them together. We denote each term by
\begin{equation}
\mathcal{G}^2_{g n_{gc} n_{gd} q n_{qc} n_{qd} \bar{q} n_{\bar{q}c} n_{\bar{q}d} }\,,
\end{equation}
where the subscripts are uniquely constructed according to the following rules:
\begin{enumerate}
\item \textbf{Particle species:} The labels $g$, $q$, and $\bar{q}$ denote the specific particle species involved in the vertex: gluons, quarks, and antiquarks, respectively.
\item \textbf{Operator counts and structure:} Following each species label, the integers $n_{xc}$ and $n_{xd}$ (where $x \in \{g,q,\bar{q} \}$) give the number of creation and annihilation operators of that species (creation first, annihilation second). For example, the term denoted by $\mathcal{G}^2_{g10 \,q 11}$ corresponds to an interaction involving exactly one gluon creation operator, a pair of quark creation and annihilation operators $\textit{plus its hermitian conjugate}$. Consequently,  $\mathcal{G}^2_{g10 \,q 11} \propto (a^\dagger b^\dagger b + \text{h.c})$.
\item \textbf{Ordering and convention:} We normal-ordered every term, and since each of them consists of a sum of an operator and its Hermitian conjugate ($\mathcal{G}^2_{\text{subscript}} + \text{h.c}$), we assign its subscripts based on the component containing the highest number of creation operators. For instance, for a term proportional to $(a^\dagger a^\dagger + aa)$, we use the label $\mathcal{G}^2_{g20}$ rather than $\mathcal{G}^2_{g02}$.
\end{enumerate}
We organize the presentation of $\mathcal{G}^2$ according to the order of the coupling constant $g$. For terms containing contributions at multiple orders, we explicitly denote the component proportional to $g^n$ using the notation $\left.\dots\right|_{g^n}$:
\begin{align}
    \cG^2|_{g^0} &= \cG^2_{0}|_{g^0}+\cG^2_{g20}|_{g^0} + \cG^2_{g11}|_{g^0} , \nonumber \\
    \cG^2|_{g^1} &= \cG^2_{g30}+\cG^2_{g21} + \cG^2_{g10q11} + \cG^2_{g10\bar{q}11}+ \cG^2_{g10q10\bar{q}10} +\cG^2_{g10q01\bar{q}01} ,\nonumber \\
    \cG^2|_{g^2} &= \cG^2_{0}|_{g^2}+\cG^2_{g20}|_{g^2} + \cG^2_{g11}|_{g^2} +\cG^2_{g40} + \cG^2_{g31} + \cG^2_{g22} + \cG^2_{g20q11} +  \cG^2_{g20\bar{q}11} + \cG^2_{g20q10\bar{q}10} + \cG^2_{g20q01\bar{q}01} + \cG^2_{g11q11}\nonumber\\
    & + \cG^2_{g11\bar{q}11} + \cG^2_{g11q10\bar{q}10} + \cG^2_{q20\bar{q}20} + \cG^2_{q21\bar{q}10} + \cG^2_{q10\bar{q}21}+\cG^2_{q22}+\cG^2_{\bar{q}22}+\cG^2_{q11\bar{q}11},
     \label{eq: G^2 as a sum of self-adjoint terms}
\end{align}
The explicit formulae for these terms are tabulated below, classified by their order in the coupling constant $g$ as indicated in the last equation (\ref{eq: G^2 as a sum of self-adjoint terms}). In the following tables, momentum integrations and momentum-conserving delta functions are omitted. (Every term is, unless the momentum integrals are specified, understood to carry an integration over all external momenta and an overall momentum-conservation Dirac's delta.) We use
\begin{equation}
\int[p_1\cdots p_n]\equiv  
\prod_{r=1}^{n}\int \frac{d^3p_r}{(2\pi)^{3/2}\sqrt{2E_{p_r}}}, \ \ \
\widetilde{\delta}(P)\equiv (2\pi)^3\delta^{(3)}(P). \quad\text{(omitted)}
\end{equation}
Momentum integrals not expressible in this form are explicitly written.
We actually list $\frac{1}{\Lambda^2}\int d^3 x \,\mathcal{G}^2$ which is the quantity (with dimensions of energy) that will be employed in the Lagrange-modified minimization of $\langle H \rangle$.

The squared Gauss' law operator terms of order $g^0$, are spelled out in Table~\ref{tab:order_g0}, those of order $g$ in Table~\ref{tab:order_g1}, involving cubic vertices and finally, those of order $g^2$ in Table~\ref{tab:order_g2},  which includes quartic vertices, products of cubic vertices and constant terms obtained from normal ordering. 

In these tables, $N$ is the number of points in the discretized momentum grid, $N_c$ is the number of colors, $N^q_s$ is the number of quark spin states, $N^g_s$ is the number of gluon polarizations and $N_f$ is the number of quark flavors

\begin{table}[h]
    \centering
\caption{Contribution to $ \int d^3 x \,\mathcal{G}^2$ from each term of order $\mathcal{O}(g^0)$.} \label{tab:order_g0}
\begin{tabular}{c||c}
\textbf{Symbol} & \textbf{Operator} \\ 
\hhline{=#=}
$\left.\mathcal{G}^2_{0}\right|_{g^0}$ & \begin{minipage}[c][1cm]{6cm} $\frac{(N_c^2-1)}{2}\, \delta^{(3)}(0)\int d^3p\, E_{p}(E_{p}^2-m_g^2)$\end{minipage}\\ \hhline{-||- }
$\left.\mathcal{G}_{g20}^{2}\right|_{g^0}$ & \begin{minipage}[c][1cm]{6cm} $ \int d^3p
\left(-\frac{E_{p}}{2}p^{i}p^{j}\right)
\left(a_{pia}^{\dagger}a_{-pja}^{\dagger}+\text{h.c.}\right)$\end{minipage}\\ \hhline{-||- }
$\left.\mathcal{G}_{g11}^{2}\right|_{g^0}$ & \begin{minipage}[c][1cm]{6cm} $\int d^3p
E_{p}p^{i}p^{j} a_{pia}^{\dagger}a_{pja}
$\end{minipage}
\end{tabular}
\end{table}

\begin{table}[h]
    \centering
\begin{minipage}{14cm}\caption{ Contribution to $\int d^3 x \,\mathcal{G}^2$ from each term of order $\mathcal{O}(g^1)$. Each particle's momentum is integrated over and global momentum conservation enforced by a $\delta(\sum p_i)$ (omitted).} \label{tab:order_g1}\end{minipage}
\begin{tabular}{c||c}
\textbf{Symbol} & \textbf{Operator} \\ 
\hhline{=#=}
$\mathcal{G}_{g30}^{2}$ & \begin{minipage}[c][1cm]{10cm} $2g f^{abc} E_{p}E_{k} p^{i} \left(ia_{pia}^{\dagger}a_{kjb}^{\dagger}a_{qjc}^{\dagger}+\text{h.c.}\right)$\end{minipage}\\ \hhline{-||- }
$\mathcal{G}_{g21}^{2}$ & \begin{minipage}[c][1cm]{10cm} $ 2g f^{abc} \Big(E_{p}E_{k}p^{i}\delta^j_l +E_{p}E_{q}q^{l}\delta^{i}_{j}-E_{k}E_{q}k^{j}\delta^i_l\Big) \left(ia_{pia}^{\dagger}a_{kjb}^{\dagger}a_{qlc}+\text{h.c.}\right)$\end{minipage}\\ \hhline{-||- }
$\mathcal{G}_{g10q11}^{2}$ & \begin{minipage}[c][1cm]{10cm} $-2g E_{p}p^{i}  \Big(T_{rr'}^{a} u_{\lambda f}^{\dagger}(k)u_{\sigma f}(q)a_{pia}^{\dagger}b_{k\lambda rf}^{\dagger}b_{q\sigma r'f}+\text{h.c.}\Big)$\end{minipage} \\  \hhline{-||- }
$\mathcal{G}_{g10\bar{q}11}^{2}$ & \begin{minipage}[c][1cm]{10cm} $2g E_{p}p^{i} \Big(T_{r'r}^{a}  v_{\sigma f}^{\dagger}(q)v_{\lambda f}(k) a_{pia}^{\dagger}d_{k\lambda rf}^{\dagger}d_{q\sigma r'f}+\text{h.c.}\Big)$\end{minipage} \\  \hhline{-||- }
$\mathcal{G}_{g10q10\bar{q}10}^{2}$ & \begin{minipage}[c][1cm]{10cm} $-2g E_{p}p^{i} \Big(T_{rr'}^{a} u_{\lambda f}^{\dagger}(k) v_{\sigma f}(q)a_{pia}^{\dagger}b_{k\lambda rf}^{\dagger}d_{q\sigma r'f}^{\dagger}+\text{h.c.}\Big)$\end{minipage} \\  \hhline{-||- }
$\mathcal{G}_{g10q01\bar{q}01}^{2}$ & \begin{minipage}[c][1cm]{10cm} $-2gE_{p}p^{i}  \Big(T_{rr'}^{a} v_{\lambda f}^{\dagger}(k)u_{\sigma f}(q)a_{pia}^{\dagger}d_{k\lambda rf}b_{q\sigma r'f}+\text{h.c.}\Big)$\end{minipage}
\end{tabular}
\end{table}

\begin{table}[]
    \centering
\begin{minipage}{16cm}\caption{Contribution to $\int d^3 x \,\mathcal{G}^2$ from each term of order $\mathcal{O}(g^2)$. If not explicitly shown, it is understood that each particle's momentum is integrated over and global momentum conservation enforced by a $\delta(\sum p_i)$ (omitted).} \label{tab:order_g2}\end{minipage}
\begin{tabular}{c||c}
\textbf{Symbol} & \textbf{Operator} \\ 
\hhline{=#=}
$\left.\mathcal{G}^2_{0}\right|_{g^2}$ & \begin{minipage}[c][1cm]{12cm} $ \frac{g^2}{2}\delta^{(3)}(0)\frac{(N_c^2-1)N_cN_s^g}{4(2\pi)^3}
\int d^3p\, d^3k\,\frac{(E_k-E_p)^2}{E_kE_p}+\frac{2g^2}{(2\pi)^3}(N_s^q)^2(N_c^2-1)(1+2N)^3\Lambda
$\end{minipage} \\  \hhline{-||- }
$\left.\mathcal{G}_{g20}^{2}\right|_{g^2}$ & \begin{minipage}[c][1cm]{12cm} $\frac{N_c g^2}{4(2\pi)^3}\int d^3p \left(\int d^3k\,\frac{E_k^2-E_p^2}{E_kE_p}\right)
\delta^{ij}
\left(a_{pia}^{\dagger}a_{-pja}^{\dagger}+\text{h.c.}\right)$\end{minipage} \\  \hhline{-||- }
$\left.\mathcal{G}_{g11}^{2}\right|_{g^2}$ & \begin{minipage}[c][1cm]{12cm} $ \int d^3p
\frac{N_c g^2}{2(2\pi)^3}\left(\int d^3k\,\frac{E_k^2+E_p^2}{E_kE_p}\right)\delta^{ij}
a_{pia}^{\dagger}a_{pja}$\end{minipage} \\  \hhline{-||- }
$\mathcal{G}_{g40}^{2}$ & \begin{minipage}[c][1cm]{12cm} $  -g^{2} f^{abc}f^{ade} E_{p}E_{q}
\left(a_{pib}^{\dagger}a_{kic}^{\dagger}
a_{qjd}^{\dagger}a_{lje}^{\dagger}+\text{h.c.}\right)$\end{minipage} \\  \hhline{-||- }
$\mathcal{G}_{g31}^{2}$ & \begin{minipage}[c][1cm]{12cm} $-g^{2} f^{abc}f^{ade}
(E_{p}-E_k)(E_{q}+E_{l})\left(a_{pib}^{\dagger}a_{kic}^{\dagger}
a_{qjd}^{\dagger}a_{lje}+\text{h.c.}\right)$\end{minipage} \\  \hhline{-||- }
$\mathcal{G}_{g22}^{2}$ & \begin{minipage}[c][1cm]{12cm} $ g^{2}
\Big\{f^{abc}f^{ade} (E_{p}E_{q}+E_{l}E_{k})
\delta^{i}_{j}\delta^{i'}_{j'} + f^{abe}f^{adc}(E_p +E_l)(E_q+E_k)\delta^{i}_{j'}\delta^{j}_{i'}
\Big\}a_{pib}^{\dagger}a_{kjc}^{\dagger}a_{qi'd}a_{lj'e}$\end{minipage} \\  \hhline{-||- }
$\mathcal{G}_{g20q11}^{2}$ & \begin{minipage}[c][1cm]{12cm} $ -2g^{2} f^{abc}  E_{p} \Big(T_{rr'}^{a} u_{\lambda f}^{\dagger}(q)u_{\sigma f}(l)\, ia_{pib}^{\dagger}a_{kic}^{\dagger}b_{q\lambda r f}^{\dagger}b_{l\sigma r'f}+\text{h.c.}\Big)$\end{minipage}\\  \hhline{-||- }
$\mathcal{G}_{g20\bar{q}11}^{2}$ & \begin{minipage}[c][1cm]{12cm} $2g^{2} f^{abc}  E_{p} \Big(T_{r'r}^{a} v_{\sigma f}^{\dagger}(l)v_{\lambda f}(q) \, ia_{pib}^{\dagger}a_{kic}^{\dagger}d_{q\lambda r f}^{\dagger}d_{l\sigma r'f}+\text{h.c.}\Big)$\end{minipage}\\  \hhline{-||- }
$\mathcal{G}_{g20q10\bar{q}10}^{2}$ & \begin{minipage}[c][1cm]{12cm} $-2g^{2} f^{abc}  E_{p}\Big(T_{rr'}^{a} u_{\lambda f}^{\dagger}(q)v_{\sigma f}(l)\, ia_{pib}^{\dagger}a_{kic}^{\dagger}b_{q\lambda rf}^{\dagger}d_{l\sigma r'f}^{\dagger}
+\text{h.c.}\Big)$\end{minipage}\\  \hhline{-||- }
$\mathcal{G}_{g20q01\bar{q}01}^{2}$ & \begin{minipage}[c][1cm]{12cm} $ -2g^{2} f^{abc}  E_{p} \Big(T_{rr'}^{a} v_{\lambda f}^{\dagger}(q)u_{\sigma f}(l) \, ia_{pib}^{\dagger}a_{kic}^{\dagger}d_{q\lambda r f}b_{l\sigma r'f}+\text{h.c.}\Big)$\end{minipage}\\  \hhline{-||- }
$\mathcal{G}_{g11q11}^{2}$ & \begin{minipage}[c][1cm]{12cm} $ -2g^{2} f^{abc} T_{rr'}^{a} (E_{p}+E_{l}) u_{\lambda f}^{\dagger}(k)u_{\sigma f}(q) \, i\,a_{pib}^{\dagger}b_{k\lambda rf}^{\dagger}b_{q\sigma r'f}a_{lic}$\end{minipage}\\  \hhline{-||- }
$\mathcal{G}_{g11\bar{q}11}^{2}$ & \begin{minipage}[c][1cm]{12cm} $2g^{2} f^{abc} T_{r'r}^{a} (E_{p}+E_{l}) v_{\sigma f}^{\dagger}(q)v_{\lambda f}(k) \, i\,a_{pib}^{\dagger}d_{k\lambda rf}^{\dagger}d_{q\sigma r'f}a_{lic}$\end{minipage}\\  \hhline{-||- }
$\mathcal{G}_{g11q10\bar{q}10}^{2}$ & \begin{minipage}[c][1cm]{12cm} $-2g^{2} f^{abc}  (E_{p}+E_l) \Big(T_{rr'}^{a} u_{\lambda f}^{\dagger}(k)v_{\sigma f}(q) \, i \,a_{pib}^{\dagger}b_{k\lambda r f}^{\dagger}d^{\dagger}_{q\sigma r'f}a_{lic}+\text{h.c.}\Big)$\end{minipage}\\  \hhline{-||- }
$\mathcal{G}_{q21\bar{q}10}^{2}$ & \begin{minipage}[c][1cm]{12cm} $2g^{2}  \Big(T_{bc}^{a}T_{de}^{a} u_{\lambda f}^{\dagger}(p)u_{\sigma f}(k) u_{\eta m}^{\dagger}(q) \, v_{\xi m}(l) b_{p\lambda bf}^{\dagger}b_{q\eta dm}^{\dagger}d_{l\xi em}^{\dagger}b_{k\sigma cf} +\text{h.c.}\Big)$\end{minipage}\\  \hhline{-||- }
$\mathcal{G}_{q10\bar{q}21}^{2}$ & \begin{minipage}[c][1cm]{12cm} $2g^{2}  \Big( T_{bc}^{a}T_{de}^{a}u_{\lambda f}^{\dagger}(p)v_{\sigma f}(k) v_{\eta m}^{\dagger}(q) \, v_{\xi m}(l) b_{p\lambda bf}^{\dagger}d_{l\xi em}^{\dagger}d_{k\sigma cf}^{\dagger}d_{q\eta dm} +\text{h.c.}\Big)$\end{minipage}\\  \hhline{-||- }
$\mathcal{G}_{q11\bar{q}11}^{2}$ & \begin{minipage}[c][1.5cm]{12cm} $ 2g^{2} \Big\{ \delta_{fm}\delta_{sr}T_{bc}^{a}T_{ed}^{a} \, u_{\lambda f}^{\dagger}(p)v_{\sigma m}(k)v_{\xi r}^{\dagger}(l)u_{\eta s}(q) +\delta_{fs}\delta_{mr} T_{bd}^{a}T_{ec}^{a}\, u_{\lambda f}^{\dagger}(p)u_{\eta s}(q)v_{\xi r}^{\dagger}(l)v_{\sigma m}(k)\Big\}\, b_{p\lambda bf}^{\dagger}d_{k\sigma cm}^{\dagger}d_{l\xi er}b_{q\eta ds}$\end{minipage}\\  \hhline{-||- }
$\mathcal{G}_{q20\bar{q}20}^{2}$ & \begin{minipage}[c][1cm]{12cm} $g^{2}  \Big(T_{bc}^{a}T_{de}^{a} u_{\lambda f}^{\dagger}(p)v_{\sigma f}(k) u_{\eta m}^{\dagger}(q)\, v_{\xi m}(l) b_{q\eta dm}^{\dagger}b_{p\lambda bf}^{\dagger}d_{k\sigma cf}^{\dagger}d_{l\xi em}^{\dagger} +\text{h.c.}\Big)$\end{minipage}\\  \hhline{-||- }
$\mathcal{G}_{q22}^{2}$ & \begin{minipage}[c][1cm]{12cm} $ g^{2} T_{bc}^{a}T_{de}^{a} u_{\lambda f}^{\dagger}(p)u_{\sigma f}(k) u_{\eta m}^{\dagger}(q)u_{\xi m}(l) \, b_{q\eta dm}^{\dagger}b_{p\lambda bf}^{\dagger}b_{k\sigma cf}b_{l\xi em}$\end{minipage}\\  \hhline{-||- }
$\mathcal{G}_{\bar{q}22}^{2}$ & \begin{minipage}[c][1cm]{12cm} $g^{2} T_{bc}^{a}T_{de}^{a} v_{\lambda f}^{\dagger}(p)v_{\sigma f}(k) v_{\eta m}^{\dagger}(q)v_{\xi m}(l) \, d_{l\xi em}^{\dagger}d_{k\sigma cf}^{\dagger}d_{p\lambda bf}d_{q\eta dm}$\end{minipage}
\end{tabular}
\end{table}

\section{Complete expansion of the Hamiltonian in time-axial gauge}
\label{Apendix:Hamiltonian-expansion}
Now we proceed to spell out in full the Hamiltonian operator in Weyl's time-axial gauge, also in terms of the same momentum-space particle creation and destruction operators, with the same indices and momentum-integral notation as in sec.~\ref{Apendix:Gauss}.

As with the Gauss operator, we decompose the Hamiltonian in powers of the coupling constant $g$. 
\begin{align}
    H|_{g^0} & = H_0 + H_{g20} + H_{g11} + H_{q11} + H_{\bar{q}11},\nonumber\\
    H|_{g^1} & =  H_{g30} + H_{g21} + H_{g10q11} + H_{g10\bar{q}11} + H_{g10q10\bar{q}10} + H_{g10q01\bar{q}01},\nonumber \\
    H|_{g^2} &= H_{g40} + H_{g31} + H_{g22} ,
\end{align}
With this decomposition, formulae is organized in tables \ref{tab:H_order_g0}, \ref{tab:H_order_g1} and \ref{tab:H_order_g2}, corresponding to the contributions of orders  $\mathcal{O}(g^0)$, $\mathcal{O}(g^1)$ and $\mathcal{O}(g^2)$.  The factor 
\begin{equation}
M_g^2 = g^2N^q_c(N^g_s-1)\int\frac{d^3k}{2(2\pi)^3E_k}
\end{equation}
that appears in $H_0$, $H_{g11}$ and $H_{g20}$ originates from the normal ordering of the four-gluon field interaction vertex.

\begin{table}[h]
    \centering
\caption{Contribution to $H_{\rm QCD}$ from each term of order $\mathcal{O}(g^0)$.} \label{tab:H_order_g0}
\begin{tabular}{c||c}
\textbf{Symbol} & \textbf{Operator} \\ 
\hhline{=#=}
${H}_{0}$ & \begin{minipage}[c][1cm]{8cm} $\frac{N_c^2-1}{4}\,
\delta^{(3)}(0)\int \frac{d^3p}{2E_p}\Big[2E_p^2(2N_s^g-1)+N_s^g M_g^2-2m_g^2(N_s^g-1)\Big]$\end{minipage}\\ \hhline{-||- }
${H}_{g20}$ & \begin{minipage}[c][1cm]{8cm} $ \int \frac{d^3p}{4E_p}
\Big[\delta^{ij}(M_g^2-m_g^2)-p^ip^j\Big]\left(a_{pia}^{\dagger}a_{-pja}^{\dagger}+\text{h.c.}\right)$\end{minipage}\\ \hhline{-||- }
${H}_{g11}$ & \begin{minipage}[c][1cm]{8cm} $\int \frac{d^3p}{2E_p}
\Big[(2E_p^2+M_g^2-m_g^2)\delta^{ij}-p^ip^j\Big]a_{pia}^{\dagger}a_{pja}$\end{minipage}\\ \hhline{-||- }
${H}_{q11}$ & \begin{minipage}[c][1cm]{8cm} $\int d^3p\, E_{\vec p}\,
b_{p\lambda f c}^{\dagger}b_{p\lambda f c}$\end{minipage}\\ \hhline{-||- }
${H}_{\bar{q}11}$ & \begin{minipage}[c][1cm]{8cm} $\int d^3p\, E_{\vec p}\,
d_{p\lambda f c}^{\dagger}d_{p\lambda f c}$\end{minipage}
\end{tabular}
\end{table}

\begin{table}[h]
    \centering
\begin{minipage}{12cm}\caption{Contribution to $H_{\rm QCD}$ from each term of order $\mathcal{O}(g^1)$. Each particle's momentum is integrated over and the corresponding momentum-conserving delta function is omitted.}\label{tab:H_order_g1}\end{minipage}
\begin{tabular}{c||c}
\textbf{Symbol} & \textbf{Operator} \\ 
\hhline{=#=}
${H}_{g30}$ & \begin{minipage}[c][1cm]{8cm} $g f^{abc}
\left(i q^i a_{pia}^{\dagger}a_{kjb}^{\dagger}a_{qjc}^{\dagger}+\text{h.c.}\right)$\end{minipage}\\ \hhline{-||- }
${H}_{g21}$ & \begin{minipage}[c][1cm]{8cm} $ g f^{abc}
\left(p^j\delta^i_m+k^m\delta^i_j+q^j\delta^i_m
\right)\,\left(i a_{pia}^{\dagger}a_{kjb}^{\dagger}a_{qmc}
+\text{h.c.}\right)$\end{minipage}\\ \hhline{-||- }
${H}_{g10q11}$ & \begin{minipage}[c][1cm]{8cm} $ g 
\Big(T^a_{cc'}\,u_{\lambda f}^{\dagger}(k)\alpha^i u_{\sigma f}(q)a_{pia}^{\dagger}b_{k\lambda c f}^{\dagger}b_{q\sigma c'f}+\text{h.c.}\Big)$\end{minipage}\\ \hhline{-||- }
${H}_{g10\bar{q}11}$ & \begin{minipage}[c][1cm]{8cm} $ -g \Big(T^a_{c'c}
\,v_{\sigma f}^{\dagger}(q)\alpha^i v_{\lambda f}(k)a_{pia}^{\dagger}d_{k\lambda c f}^{\dagger}d_{q\sigma c'f}+\text{h.c.}\Big)$\end{minipage}\\ \hhline{-||- }
${H}_{g10q10\bar q10}$ & \begin{minipage}[c][1cm]{8cm} $ g \Big(T^a_{cc'}
\,u_{\lambda f}^{\dagger}(k)\alpha^i v_{\sigma f}(q)a_{pia}^{\dagger}b_{k\lambda c f}^{\dagger}d_{q\sigma c'f}^{\dagger}+\text{h.c.}\Big)$\end{minipage}\\ \hhline{-||- }
${H}_{g10q01\bar q01}$ & \begin{minipage}[c][1cm]{8cm} $ g \Big(T^a_{cc'}
\,v_{\lambda f}^{\dagger}(k)\alpha^i u_{\sigma f}(q)a_{pia}^{\dagger}d_{k\lambda c f}b_{q\sigma c'f}+\text{h.c.}\Big)$\end{minipage}
\end{tabular}
\end{table}

\begin{table}[h]
    \centering
\begin{minipage}{14cm}\caption{Contribution to $H_{\rm QCD}$ from each term of order $\mathcal{O}(g^2)$. Each particle's momentum is integrated over and the corresponding momentum-conserving delta function is omitted.} \label{tab:H_order_g2}\end{minipage}
\begin{tabular}{c||c}
\textbf{Symbol} & \textbf{Operator} \\ 
\hhline{=#=}
${H}_{g40}$ & \begin{minipage}[c][1cm]{10cm} $\frac{g^2}{4} f^{abc}f^{ade}
\left(a_{pib}^{\dagger}a_{kjc}^{\dagger}a_{qjd}^{\dagger}a_{lie}^{\dagger}
+\text{h.c.}\right)$\end{minipage}\\ \hhline{-||- }
${H}_{g31}$ & \begin{minipage}[c][1cm]{10cm} $\frac{g^2}{2}
\left(f^{abc}f^{ade}+f^{abe}f^{adc}\right)\,\left(a_{pib}^{\dagger}a_{kjc}^{\dagger}a_{qjd}^{\dagger}a_{lie}+\text{h.c.}\right)$\end{minipage}\\ \hhline{-||- }
${H}_{g22}$ & \begin{minipage}[c][1cm]{10cm} $\frac{g^2}{2}
\Big\{f^{abc}f^{ade}\delta^i_{i'}\delta^j_{j'} +f^{abe}f^{adc}\left(\delta^i_{i'}\delta^j_{j'}-\delta^i_j\delta^{j'}_{i'}\right)\Big\}\,
a_{pib}^{\dagger}a_{kjc}^{\dagger}a_{qi'd}a_{lj'e}$\end{minipage}\\ \hhline{-||- }
\end{tabular}
\end{table}
\smallskip
\twocolumngrid

\section{Spinor Basis}
\label{sec:spinor_basis}

In this section we explicitly state the definitions, normalizations, and bilinear products of the spinor fields used throughout our calculations. Spin-$1/2$ fermion fields are expanded in terms of creation/destruction operators and spinor solutions of the free Dirac equation,
\begin{equation}
u_{\vec{k}\lambda f}=\sqrt{E(|\vec{k}|, f)} 
\begin{pmatrix} 
\sqrt{1+s_{kf}}\chi_\lambda \\ 
\sqrt{1-s_{kf}}\vec{\sigma}\cdot\hat{k}\chi_\lambda 
\end{pmatrix} , \label{eq:u_spinor}
\end{equation}
while for antiparticles they read
\begin{equation}
v_{-\vec{k}\lambda f}=\sqrt{E(|\vec{k}|,f)} 
\begin{pmatrix} 
-\sqrt{1-s_{kf}}\, i\vec{\sigma}\cdot\hat{k}\sigma_2\chi_\lambda \\ 
\sqrt{1+s_{kf}}\, i\sigma_2\chi_\lambda 
\end{pmatrix} , \label{eq:v_spinor}
\end{equation}
where $\vec{\sigma}=(\sigma_x,\sigma_y,\sigma_z)$ are the standard Pauli matrices (so that $i\sigma_2$ is the appropriate to obtain charge-conjugated spinors), $\chi_\lambda$ are two-dimensional orthonormal basis vectors, and $\hat{k} = \vec{k}/|\vec{k}|$. The coefficients $s_{kf}$ and $c_{kf}$ are defined in terms of the Bogoliubov angle $\phi_{k,f}$ encoding a running mass $m(|\vec{k}|,f)$:
\begin{equation}
s_{kf}=\sin\phi_{k,f}=\frac{m(|\vec{k}|,f)}{E(|\vec{k}|,f)},\quad c_{kf}=\cos\phi_{k,f}=\frac{|\vec{k}|}{E(|\vec{k}|,f)} .
\end{equation}
We take $m(|\vec{k}|,f) = m_f$ as the bare mass, so that the angle is fixed, and $\psi_f$ is a free quark field. Consequently, the polarization functions satisfy the Dirac equations:
\begin{align}
(\gamma\cdot p -m) u_{\vec{k}\lambda' f} &= \left(\gamma^0 E_{\vec{k}} - \vec{\gamma}\cdot\vec{k} - m\right) u_{\vec{k}\lambda' f}=0, \\
(\gamma\cdot p+m) v_{-\vec{k}\lambda f} &=\left(\gamma^0 E_{\vec{k}}+\vec{\gamma}\cdot\vec{k}+m\right)v_{-\vec{k}\lambda f}=0 .
\end{align}
In the standard Pauli-Dirac representation, the $\gamma$ matrices read:
\begin{equation}
\gamma^0=\begin{pmatrix}\mathbb{I}_2 &0\\0&-\mathbb{I}_2\end{pmatrix},\quad \gamma^i=\begin{pmatrix}0 & \sigma^i \\ -\sigma^i & 0 \end{pmatrix} ,
\end{equation}
and fulfill the Dirac algebra $\{\gamma^\mu,\gamma^\nu\}=2g^{\mu\nu}$.

These spinors satisfy the following Lorentz-covariant normalization conditions:
\begin{align}
u^\dagger_{\vec{k}\lambda f} u_{\vec{k}\lambda' f} &= 2 E(|\vec{k}|,f) \delta_{\lambda\lambda'} , \\
v^\dagger_{-\vec{k}\lambda f} v_{-\vec{k}\lambda' f} &= 2 E(|\vec{k}|,f) \delta_{\lambda\lambda'} , \\
u^\dagger_{\vec{k}\lambda f} v_{-\vec{k}\lambda' f} &= v^\dagger_{-\vec{k}\lambda f} u_{\vec{k}\lambda' f} = 0 .
\end{align}

The relevant covariant bilinear products evaluated with the normalizations above yield:
\begin{align}
\bar{u}_{\vec{k}\lambda f} u_{\vec{k}\lambda' f} &= 2 m(|\vec{k}|,f) \delta_{\lambda\lambda'} , \\
\bar{v}_{-\vec{k}\lambda f} v_{-\vec{k}\lambda' f} &= -2 m(|\vec{k}|,f) \delta_{\lambda\lambda'} , \\
\bar{u}_{\vec{k}\lambda f} v_{-\vec{k}\sigma f} &= 2 \chi_\lambda^\dagger \left(k^1\sigma^3-ik^2 \mathbb{I}_2-k^3 \sigma^1\right)\chi_\sigma .
\end{align}

Furthermore, the bilinear products involving $\alpha^i=\gamma^0 \gamma^i$ yield:
\begin{align}
u^\dagger_{\vec{k}\lambda f}\gamma^0\gamma^i u_{\vec{k}\lambda' f} &= 2k^i\delta_{\lambda\lambda'} , \\
v^\dagger_{-\vec{k}\lambda f}\gamma^0\gamma^i v_{-\vec{k}\lambda' f} &= -2k^i\delta_{\lambda\lambda'} .
\end{align}

Finally, the polarization sums (outer products) evaluate to the following projectors:
\begin{align}
\sum_{\lambda}u_{\vec{k}\lambda f}u_{\vec{k}\lambda f}^\dagger &= \left(\slashed{k}+m_f(|\vec{k}|)\right)\gamma^0 , \\
\sum_{\lambda}v_{\vec{k}\lambda f}v_{\vec{k}\lambda f}^\dagger &= \left(\slashed{k}-m_f(|\vec{k}|)\right)\gamma^0 ,
\end{align}
where $\slashed{k}=k^0\gamma^0-\vec{k}\cdot\vec{\gamma}$.



\begin{figure*}
    \centering
    \includegraphics[width=1\linewidth]{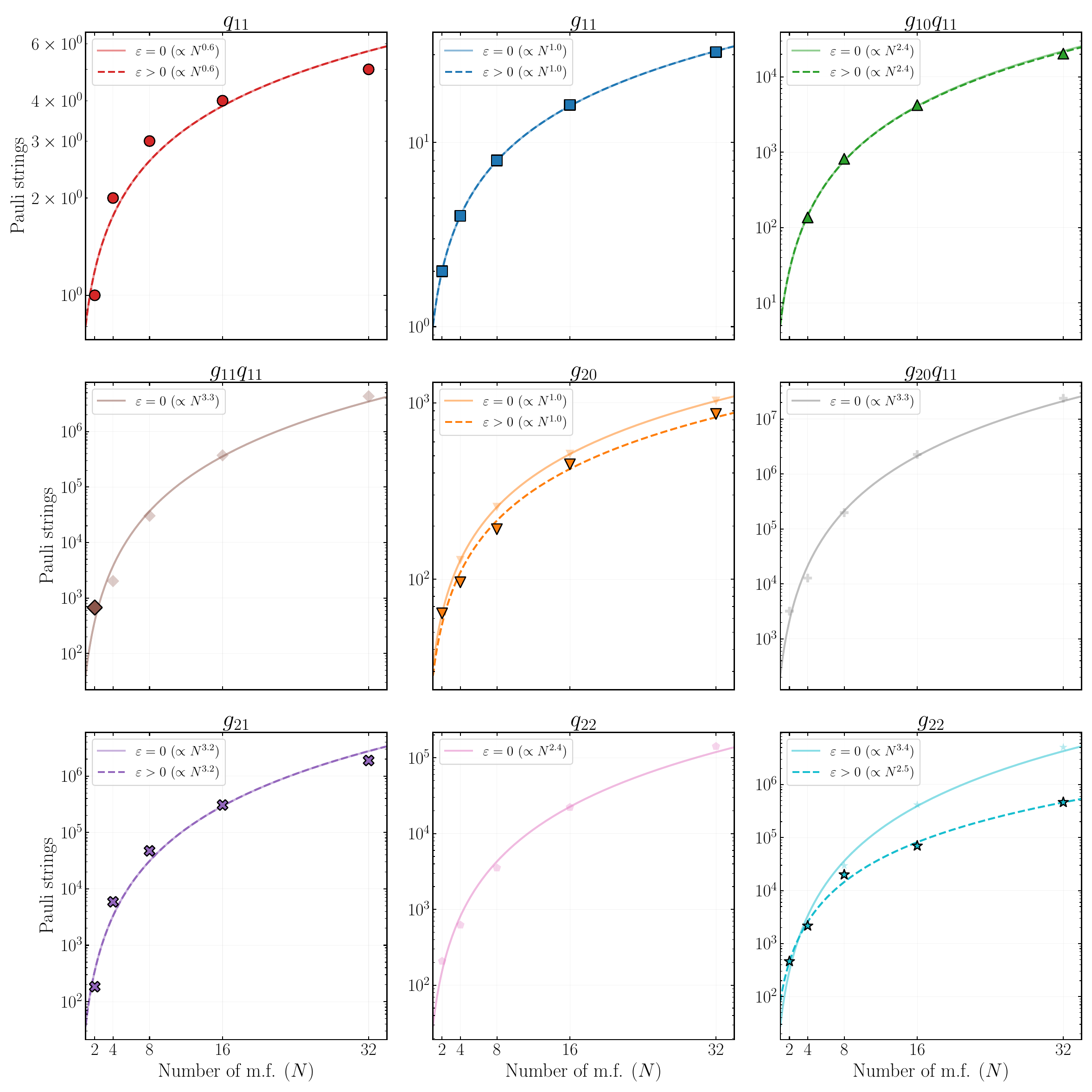}
    \caption{Number of Pauli strings for individual, selected matrix elements of the operator $H+G^2$ before and after applying the cuttof discussed in \ref{scaling}, as function of the possible number of momenta in which the grid is fractioned.}
    \label{fig:scaling por module}
\end{figure*}

\section{Alternative measurement strategy: Gray code + Hadamard test}
\label{Apendix:Gray-code}
The Gray codes provide an efficient alternative strategy for implementing unitary operators that connect two computational-basis states and for measuring their expected values through the Hadamard test. This approach is particularly useful for sparse operators, especially when one wishes to avoid a Pauli-string decomposition, which can be costly in terms of gate count~\cite{di_matteo_improving_2021} (for example, figure~\ref{fig:scaling por module} details the scaling for several matrix elements involving only gluons).

The Hadamard test enables the extraction of the real and imaginary parts of the expectation $\bra{\Psi}\mathcal{U}\ket{\Psi}$ by introducing two ancillary qubits and performing suitable controlled operations. One of these qubits is prepared in the state $\left(\ket{0}+\ket{1}\right)/\sqrt{2}$ with a Hadamard gate. A controlled version of $\mathcal{U}$ is then applied, such that $\mathcal{U}$ acts only when the first ancillary qubit is in the state $\ket{1}$. Afterward, a second Hadamard gate is applied on the same qubit, which is measured. The difference between the probability of obtaining $\ket{0}$ and $\ket{1}$ yields the real part of the expectation value. If a phase gate $\texttt{S}^\dagger=\textrm{diag}(1,-i)$ is applied to the first auxiliary qubit before the second Hadamard gate, the corresponding yields the imaginary part. Therefore,
\begin{equation}
    \bra{\Psi}\mathcal{U}\ket{\Psi}=\left(\mathcal{P}_{\ket{0}}^{\texttt{H}}-\mathcal{P}_{\ket{1}}^{\texttt{H}}\right)+i\left(\mathcal{P}_{\ket{0}}^{\texttt{S}^\dagger\to \texttt{H}}-\mathcal{P}_{\ket{1}}^{\texttt{S}^\dagger\to \texttt{H}}\right)
\end{equation}
where $\mathcal{P}^\texttt{G}_{\ket{i}}$ denotes the probability of measuring the ancillary qubit in the state $\ket{i}$ after applying the gate sequence $\texttt{G}$.

Given two states $\ket{i}$ and $\ket{j}$ defined on $n$ qubits, we define their Hamming distance, $d_H(i,j)$, as the number of bit flips required to transform $\ket{i}$ into $\ket{j}$. A Gray code consists of an ordered sequence of binary numbers where each consecutive pair differs by exactly one qubit. The Gray-code path connecting $\ket{i}$ and $\ket{j}$ is therefore a sequence of intermediate states $\ket{g_0}=\ket{i},\ket{g_1},\dots,\ket{g_{m-1}},\ket{g_m}=\ket{j}$ such that, for every $k$, $d_H(g_{k-1},g_k)=1$. For example, for $n=3$, $\ket{i}=\ket{101}$ and $\ket{j}=\ket{010}$, a valid Gray path is:
\begin{equation*}
    \ket{101}\to\ket{100}\to\ket{110}\to\ket{010}\, .
\end{equation*}

\begin{figure}[h!]
    \centering
    \begin{quantikz}[thin lines] 
         & \gate[3]{G} & \ctrl{2} &  \qw  & \qw  & \qw & \qw \\
         \lstick{$\ket{\Psi}$} & \qw  & \qw  & \gate[1]{\mathcal{U}} & \qw & \qw & \qw \\   
         & & \octrl{1} &  \qw  & \qw & \qw & \qw \\  
         \lstick{$\ket{0}$} & \gate{\texttt{H}} & \ctrl{1} & \qw & \gate[style={dashed}]{\texttt{S}^\dagger} & \gate{\texttt{H}} & \meter{} \\
         \lstick{$\ket{0}$} & \qw & \gate{\texttt{X}} &  \ctrl{-3}    & \qw & \qw & \qw
    \end{quantikz} \par
    \caption{\small Quantum circuit implementing the Hadamard test combined with the Gray-code construction. The operators $G$ and $G^\dagger$ map the computational-basis states $\ket{i}$ to the final Gray-code predecessor $\ket{j}$ state, while the controlled operation $\mathcal{U}$ acts on the two-dimensional subspace spanned by $\ket{i}$ and $\ket{j}$. Measurements of the ancillary qubits $\ket{0}$ yield the real/imaginary part of the corresponding matrix element and control the central Pauli \texttt{X}/\texttt{Y} gate from the Gray's code.}
    \label{fig:Hadamard+Gray}
\end{figure}
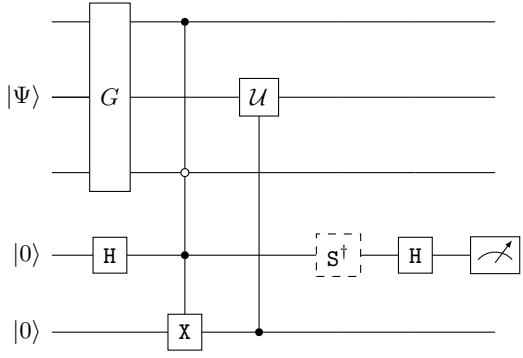

The implementation now proceeds as in Fig.(\ref{fig:Hadamard+Gray}). First, one follows the Gray-code path by applying multi-controlled \texttt{NOT} (\texttt{MCX}) that transform each state $\ket{g_{k-1}}$ into $\ket{g_k}$, where, for each transition, the target qubit is chosen as the one that changes between the two Gray-code words. The remaining qubits act as controls and are fixed according to the bit values appearing in $\ket{g_{k-1}}$. Positive and negative controls are introduced whenever a control qubit must be fixed to $1$ or $0$, respectively. For example, for the transition $\ket{101}\to\ket{100}$, qubit $0$ is selected as the target, while qubits $1$ and $2$ are fixed to the values appearing in the state $\ket{101}$. 

Once the state $\ket{g_{m-1}}$ (in the example, $\ket{110}$) is reached, the Hadamard test starts by preparing the first auxiliary qubit in a superposition. The unitary gate $U$ that is going to be measured is,
\begin{equation}
    \mathcal{U}=\alpha\ket{i}\bra{j}+\alpha^*\ket{j}\bra{i}+\left(\mathbb{I}-\ket{i}\bra{i}-\ket{j}\bra{j}\right)
\end{equation}
where $\ket{i}$ and $\ket{j}$ are computational basis states such that $d_H(i,j) = 1$, and $\alpha$ is a complex constant. 

Since $i$ and $j$ differ only in one qubit, $U$ can be implemented by applying on the qubit that differs (in the example, qubit $2$): either a Pauli \texttt{X} gate if $\alpha\in\mathbb{R}$, a Pauli \texttt{Y} gate if $i\alpha\in \mathbb{R}$ or a combination of the two if $\alpha\in\mathbb{C}$. These operations should be controlled by a \texttt{MCX} gates on the remaining qubits of $\ket{i}$ and $\ket{j}$ and the first auxiliary qubit being in state $1$. Instead of doing so, we initiate a second auxiliary qubit with the result of such a test, and then control $U$ on that qubit. In this way, the total number of \texttt{MCX} gates is $2d_H-1$. \cite{siwach_quantum_2022}

The Gray-code construction and the Hadamard test are thus combined. The Gray-code sequence first provides a unitary operator that reduces the Hamming distance between the initial states. The remaining transformation relating the final states is then promoted to a controlled operation on the ancillary qubit, as required by the Hadamard test. This creates the quantum interference needed to extract the desired matrix element. The resulting circuit is shown in Fig. \ref{fig:Hadamard+Gray}.

This procedure has been employed in simulations of nuclear Hamiltonians and multilevel quantum systems, demonstrating a significant reduction in circuit depth compared with more naive implementations \cite{sawaya_resource-efficient_2020}.

\section{Minimization data}
\label{mindata}
The tables in this section compile the results of the constrained minimization of the Hamiltonian for the three sets of parameters of table \ref{tab:parameters}. The two tables of each case correspond to both wavefunctions described in \ref{sec:varwaves}, tables to the left to the 1 quark - 2 gluon wavefunction, \ref{vwave:3q1g}, whereas tables to the right correspond to the 3 quark - 1 gluon wavefunction \ref{vwave:1q2g}.
\begin{table}[]
\caption{Data corresponding to minimization with the set of parameters ``core''. All units are in $\mathrm{GeV}.$}
\label{tab:dcore}
\begin{tabular}{|c||c|c|}
\hline
$\Lambda$ &  $H(g=0)$ & $H$ \\ \hhline{=#= =}
   3.1   & 37.2             & 71.176       \\
\hhline{-|| - -}
      3.11   & 37.32            & 71.317       \\
\hhline{-|| - -}
      3.12   & 37.44            & 71.457       \\
\hhline{-|| - -}
      3.13   & 37.56           & 71.597       \\
\hhline{-|| - -}
      3.14   & 37.68             & 71.737        \\
\hhline{-|| - -}
      3.15   & 37.80             & 71.878     \\
\hhline{-|| - -}
      3.16   & 37.92            & 72.019       \\
\hhline{-|| - -}
\end{tabular}
\begin{tabular}{|c||c|c|}
\hline
$\Lambda$ &  $H(g=0)$ & $H$ \\ \hhline{=#= =}
   2.12   & 25.44  & 49.160      \\
\hhline{-|| - -}
   2.13   & 25.56  & 49.302      \\
\hhline{-|| - -}
   2.14   & 25.68  & 49.444      \\
\hhline{-|| - -}
   2.15   & 25.80  & 49.586      \\
\hhline{-|| - -}
   2.16   & 25.92  & 49.728      \\
\hhline{-|| - -}
   2.17   & 26.04  & 49.871      \\
\hhline{-|| - -}
   2.18   & 26.16  & 50.014      \\
\hhline{-|| - -}
\end{tabular}
\end{table}

\begin{table}[]
\caption{Data corresponding to minimization with the set of parameters ``middle''. All units are in $\mathrm{GeV}.$}
\label{tab:dmid}
\begin{tabular}{|c||c|c|}
\hline
$\Lambda$ &  $H(g=0)$ & $H$ \\ \hhline{=#= =}
   4.36   & 52.32  & 89.652      \\
\hhline{-|| - -}
   4.37   & 52.44  & 89.803      \\
\hhline{-|| - -}
   4.38   & 52.56  & 89.954      \\
\hhline{-|| - -}
   4.39   & 52.68  & 90.104      \\
\hhline{-|| - -}
   4.40   & 52.80  & 90.255      \\
\hhline{-|| - -}
   4.41   & 52.92  & 90.406      \\
\hhline{-|| - -}
   4.42   & 53.04  & 90.556      \\
\hhline{-|| - -}
\end{tabular}
\begin{tabular}{|c||c|c|}
\hline
$\Lambda$ &  $H(g=0)$ & $H$ \\ \hhline{=#= =}
   3.02   & 36.24  & 62.582      \\
\hhline{-|| - -}
   3.03   & 36.36  & 62.735      \\
\hhline{-|| - -}
   3.04   & 36.48  & 62.889      \\
\hhline{-|| - -}
   3.05   & 36.60  & 63.042      \\
\hhline{-|| - -}
   3.06   & 36.72  & 63.196      \\
\hhline{-|| - -}
   3.07   & 36.84  & 63.350      \\
\hhline{-|| - -}
   3.08   & 36.96  & 63.503      \\
\hhline{-|| - -}
\end{tabular}
\end{table}

\begin{table}[]
\caption{Data corresponding to minimization with the set of parameters ``match''. All units are in $\mathrm{GeV}.$}
\label{tab:dmatch}
\begin{tabular}{|c||c|c|}
\hline
$\Lambda$ &  $H(g=0)$ & $H$ \\ \hhline{=#= =}
   5.62   & 67.44  & 108.787      \\
\hhline{-|| - -}
   5.63   & 67.56  & 108.940      \\
\hhline{-|| - -}
   5.64   & 67.68  & 109.094      \\
\hhline{-|| - -}
   5.65   & 67.80  & 109.247      \\
\hhline{-|| - -}
   5.66   & 67.92  & 109.399      \\
\hhline{-|| - -}
   5.67   & 68.04  & 109.552      \\
\hhline{-|| - -}
   5.68   & 68.16  & 109.704      \\
\hhline{-|| - -}
\end{tabular}
\begin{tabular}{|c||c|c|}
\hline
$\Lambda$ &  $H(g=0)$ & $H$ \\ \hhline{=#= =}
   3.88   & 46.56  & 75.897      \\
\hhline{-|| - -}
   3.89   & 46.68  & 76.053      \\
\hhline{-|| - -}
   3.90   & 46.80  & 76.209      \\
\hhline{-|| - -}
   3.91   & 46.92  & 76.364      \\
\hhline{-|| - -}
   3.92   & 47.04  & 76.520      \\
\hhline{-|| - -}
   3.93   & 47.16  & 76.676      \\
\hhline{-|| - -}
   3.94   & 47.28  & 76.832      \\
\hhline{-|| - -}
\end{tabular}
\end{table}


\newpage
\bibliographystyle{apsrev4-2}
\bibliography{EoSQC}
\end{document}